\documentclass[12pt]{article}

\usepackage{arydshln}

\usepackage[english]{babel}
\usepackage{setspace} % line spacing
\usepackage[letterpaper,top=2cm,bottom=2cm,left=3cm,right=3cm,marginparwidth=1.75cm]{geometry}

\usepackage{graphicx}
\usepackage{tabularx}

\usepackage[colorlinks=true, allcolors=blue]{hyperref}  % hyperlinks

\usepackage[utf8]{inputenc} % allow utf-8 input
\usepackage[T1]{fontenc}    % use 8-bit T1 fonts

\usepackage{url}            % simple URL typesetting
\usepackage{booktabs}       % professional-quality tables
\usepackage{amsfonts}       % blackboard math symbols
\usepackage{nicefrac}       % compact symbols for 1/2, etc.
\usepackage{microtype}      % microtypography
\usepackage{xcolor}         % colors
\newcommand{\xdocref}[1]{\texorpdfstring{\textcolor{blue}{#1}}{#1}} % arXiv cross-document reference display
\usepackage[authoryear]{natbib}
\usepackage{authblk}

\usepackage{amsmath,mathtools}
\usepackage{enumitem} 
\usepackage{amsthm}         %Thm, proof 
\usepackage{caption}        %subfigure
\usepackage{subcaption}     %subfigure
\usepackage[linesnumbered,ruled,vlined]{algorithm2e} % algorithm

\SetCommentSty{mycommfont}
\SetKwInput{KwInput}{Input}                % Set the Input
\SetKwInput{KwOutput}{Output}              % set the Output

\usepackage{comment}        % begin comment
\usepackage{multirow}
\usepackage{bm}
\usepackage{tikz}
\usetikzlibrary{arrows.meta}
\usepackage{tikz-3dplot}
\usepackage{booktabs}

\newtheorem{thm}{Theorem}

\newtheorem{lem}[thm]{Lemma}

\newtheorem{Assumption}{Assumption}

\newcommand{\abs}[1]{\left\vert#1\right\vert}

\def\uv{\mathbf u}
\def\vv{\mathbf v}
\def\wv{\mathbf w}

\def\yv{\mathbf y}

\def\Av{\mathbf A}

\def\Dv{\mathbf D}
\def\Ev{\mathbf E}

\def\Iv{\mathbf I}
\def\Jv{\mathbf J}

\def\Pv{\mathbf P}

\def\Sv{\mathbf S}

\def\Uv{\mathbf U}
\def\Vv{\mathbf V}

\def\Xv{\mathbf X}

\newcommand{\lambdav}{\mbox{\boldmath{$\lambda$}}}
\newcommand{\muv}{\mbox{\boldmath{$\mu$}}}

\newcommand{\Jc}{\mathcal{J}}

\newcommand{\Vc}{\mathcal{V}}

\newcommand{\Pb}{\mathbb{P}}

\newcommand{\Rb}{\mathbb{R}}

\def\e{\mathrm e}

\def\1v{\mathbf 1}
\def\0v{\mathbf 0}

\title{\bf Testing and segmentation of joint and individual components in integrative multi-source factor models}

\author[1]{Kyoowon Kim}
\author[2]{Sungkyu Jung}

\affil[1]{Department of Statistics, Seoul National University, kwkim1224@snu.ac.kr}
\affil[2]{Department of Statistics and Institute for Data Innovation in Science, Seoul National University, sungkyu@snu.ac.kr}
\date{June 1, 2026}

\begin{document}

\maketitle

\begin{spacing}{1}
\begin{abstract}
    Disentangling shared (joint) structures from source-specific (individual) variations is a fundamental task in multi-source data integration. Existing joint–individual models often rely on computationally intensive optimization or loose spectral bounds, leading to suboptimal separation accuracy and poor scalability. In this paper, we propose the Multi-Source Sequential Alignment Test (MSSAT). MSSAT leverages the geometric observation that true joint components manifest as closely aligned score subspaces across different data sources. By deriving the asymptotic null distribution of our alignment statistic, we develop a rigorous, resampling-free sequential testing procedure to accurately estimate the joint rank. Extensive simulations and real data applications, including a TCGA multi-omics dataset, demonstrate that MSSAT achieves superior separation accuracy and substantially faster computation compared to competing methods.
\end{abstract}
\end{spacing}


\section{Introduction}
Multi-source data, also referred to as multi-view or multi-block data, comprise various types of measurements collected from a common set of samples. A prominent example is the breast cancer dataset from The Cancer Genome Atlas \citep{cancer2008comprehensive}, which integrates gene expression, copy number variation, reverse-phase protein array, and mutation data from the same subjects. Each data source may exhibit its own specific patterns or signals while simultaneously containing components shared across sources. Separating the shared patterns across multiple sources (the \emph{joint structure}) from those unique to each source (the \emph{individual structure}) allows one to understand both the common underlying effects and the source-specific variations inherent in the data.

To separate these structures, extensions of traditional CCA-based approaches such as \cite{zhou2015group, murden2022interpretive, shu2020d} have been proposed for multi-source integration. 
Because these methods are generally restricted to identifying only joint patterns among data sources, many subsequent works are based on integrative multi-source factor models that explicitly decompose the data matrix into joint and individual components. These models characterize a joint component by sharing common factor scores, while the scores corresponding to individual components are specific to one of the data sources. Building on this, several works adopt factor-model-based estimation approaches to identify latent factors either shared across sources or specific to individual sources \citep{shen2009integrative, virtanen2012bayesian, li2017incorporating, argelaguet2018multi, gao2021covariate}.
 
Alternatively, a majority of previous studies consider deterministic joint--individual components \citep{lock2013joint, feng2018angle, gaynanova2019structural, chen2022two, lock2022sjive, wang2024joint, yang2025estimating, prothero2024data, sergazinov2025spectral, park2020integrative, lock2022bidimensional}. 
While optimization-based estimators for these models \citep{lock2013joint, gaynanova2019structural, park2020integrative, lock2022bidimensional} are conceptually appealing, solving the associated non-convex problems is computationally prohibitive for large datasets. To circumvent this, methods such as AJIVE \citep{feng2018angle} and DIVAS \citep{prothero2024data} estimate the joint rank by thresholding the singular values of the concatenated data matrix. This relies on the geometric intuition that principal angles between shared subspaces should be small, which leads to large singular values. However, their perturbation-based lower bounds are typically overly conservative, frequently leading to rank overestimation. Moreover, computing these lower bounds relies on resampling procedures such as the bootstrap, which can be computationally slow. Recent theoretical work has identified the fundamental limits of such angle-based methods \citep{yang2025estimating}, and while new approaches such as the product of projections have been proposed \citep[][PPD]{sergazinov2025spectral}, they still rely on time-consuming resampling-based procedures.

To address these limitations, we propose the Multi-Source Sequential Alignment Test (MSSAT), a powerful and computationally efficient method for joint–individual decomposition in multi-source data. While MSSAT builds upon the geometric intuition of existing angle-based methods \citep{feng2018angle, prothero2024data, sergazinov2025spectral}---namely, that closely aligned score subspaces across different data sources indicate a shared joint component---our key contribution to the literature is to formally frame the task of joint rank estimation as a sequential hypothesis testing problem. Specifically, we develop closed-form testing procedures that operate rigorously under high-dimensional asymptotic regimes where $n, p \to \infty$. As a result, the proposed MSSAT procedure is completely resampling-free and thus computationally fast, while demonstrating high accuracy in decomposing joint and individual components from noisy data. To achieve this, the proposed MSSAT framework consists of three main steps: (i) extracting the low-rank signal structures from each data source, (ii) estimating the joint rank by sequentially testing for the presence of shared structures across blocks, and (iii) reconstructing the final joint and individual components for each data block.

The rest of the paper is organized as follows. Section~\ref{sec:model} introduces the integrative multi-source factor model, and Section~\ref{sec:theory} establishes the theoretical foundation for the alignment tests. Section~\ref{sec:method} details the proposed MSSAT framework. Section~\ref{sec:numerical} presents comprehensive simulation studies comparing MSSAT with existing methods, including AJIVE \citep{feng2018angle}, DIVAS \citep{prothero2024data}, BIDIFAC+ \citep{park2020integrative, lock2022bidimensional}, and PPD \citep{sergazinov2025spectral}. These numerical experiments demonstrate that MSSAT achieves superior estimation accuracy for the ranks of joint components while computationally efficient. 
Finally, Section~\ref{sec:realdata} illustrates the practical utility of our approach through applications to real-world datasets, including the TCGA breast cancer multi-omics dataset and Korean mortality data. 
The Supplementary Material contains further technical details and additional numerical results. MATLAB and R implementations of the proposed method are available at \url{https://github.com/qone1234-10/MSSAT}.

\section{Integrative multi-source factor model}\label{sec:model}

Consider a multi-source dataset comprising $K$ data blocks, where each block ${\Xv}_k\in\mathbb{R}^{p_k\times n}$ contains $p_k$ variables measured across $n$ common subjects. 
For each block, we assume a noise-additive model:
$$
\Xv_k = \muv_k + \Av_k + \Ev_k,
$$
where $\Av_k$ is a low-rank signal matrix, and $\Ev_k$ denotes the noise components. Throughout this paper, we assume $\muv_k = \0v$ and that the data are pre-centered. 

The integrative multi-source factor model further decomposes each signal matrix $\Av_k$ into two parts: a joint component $\Jv_k$ representing the common factors across different blocks, and an individual component $\Iv_k$, capturing source-specific factors. This leads to the following decomposition: 
\begin{equation}
    \label{model:jive}
    \Xv_k = \Jv_k + \Iv_k + \Ev_k,
\end{equation}
in which the joint components $\Jv_k$ share common factors across sources, while the individual components $\Iv_k$ remain source-specific. Formally, we impose the following conditions:
\begin{equation}
\begin{aligned}
    \text{row}(\Jv_1) &= \dots = \text{row}(\Jv_K), \quad \bigcap_{k=1}^{K} \text{row}(\Iv_k) = { \vec{0} }, \\
    \text{row}(\Jv_k) &\perp \text{row}(\Iv_k) \quad \mbox{for all }\,  k \in [K], 
    \end{aligned}
    \label{eq:joint_subspace}
\end{equation}
where $[K] = \{1,\ldots, K\}$. 
These conditions ensure identifiability of the joint and individual components \citep{lock2013joint,feng2018angle}. 
This framework of integrative factor model has been widely used to explore and identify shared variations among multi-source data \citep{ gaynanova2019structural, park2020integrative, lock2022bidimensional, chen2022two, prothero2024data, yang2025estimating, sergazinov2025spectral}.

The noise-additive model, where each signal matrix $\Av_k$ is deterministic with an unknown rank $r_k$ \citep{gavish2014optimal,bao2021singular}, is assumed to facilitate theoretical developments. 
We denote the joint rank as $r_{\textsc{j}}$, which implies that the individual rank for the $k$-th block is $r_k - r_{\textsc{j}}$. The noise matrices $\Ev_k$ are mutually independent across blocks, with entries $(\Ev_k)_{ij} \overset{iid}{\sim} N\left( 0, \sigma_k^2 \right)$  for $i \in [p_k]$, $j \in [n]$. 

\section{Theoretical foundation of alignment testing}\label{sec:theory}
This section establishes the theoretical foundations for measuring the geometric alignment between row subspaces. Given that  the joint components $\Jv_k$ share a common row subspace (\ref{eq:joint_subspace}), we characterize the asymptotic behavior of singular vector alignment under a high-dimensional regime for both two-block and multi-block configurations. These distributional results provide the statistical framework for joint-rank estimation procedure introduced in Section \ref{subsec:seq_align_test}. 

Throughout, we assume the following. Consider the singular value decomposition of the signal matrix 
\begin{equation}
    \label{eq:signal_svd}
    \Av_k = \Uv_{(k)}\Dv_{(k)}\Vv_{(k)}^\top = \sum_{l=1}^{r_k} d_{(k),l}\uv_{(k),l}\vv_{(k),l}^\top,
\end{equation}
and write $\lambda_{(k),l} := \lim_{n,p\to\infty}{d_{(k),l}}/({\sqrt{n}\sigma_k})$ for the limiting signal-to-noise ratio (SNR). 

\begin{Assumption}(high-dimensional regime and supercritical condition)\label{assump:high-dim_supercritical}    
    \begin{enumerate}
        \item High-dimensional regime: For each $k \in [K]$, $p_k \equiv p_k(n)$ diverges and $y_k:= p_k/n$ converges to a constant $\tilde{y}_k \in [\tau, \tau^{-1}]$ as $n \to \infty$ for some small constant $\tau \in (0,1)$.
        
        \item Supercritical condition: There exist constants $C>0$ and $\delta >0$, such that, for all $k$ and $l$, 
        $\lambda_{(k),l} \in (y_k^{1/4}+\delta ,C)$
        and $| \lambda_{(k),l+1}-\lambda_{(k),l} | \geq \delta$.
    \end{enumerate}
\end{Assumption}

Assumption~\ref{assump:high-dim_supercritical} accommodates both low and high-dimensional situations. Throughout, we assume $y_k = p_k/n = \tilde{y}_k$ for all $n$ for simplicity. The supercritical condition assumes that the SNRs, $\lambda_{(k),l}$, are  sufficiently strong and well-separated. The quantity $y_k^{1/4}$ in the lower bound corresponds to the Baik--Ben Arous--Péché phase transition boundary, which separates detectable and undetectable spikes~\citep{baik2005phase}. Both conditions are standard in the random matrix theory literature \citep{bao2021singular,benaych2012singular,paul2007asymptotics}.
  
\subsection{Asymptotic alignment in the two-block case}\label{subsubsec:two_block}
 
For $K=2$ data blocks, we develop a hypothesis test for the alignment of right-singular vectors. 
Let $\vv_{(k),\ell_k}$ denote the $\ell_k$-th right-singular vector of block $k$ for $\ell_k \in [r_k]$ from \eqref{eq:signal_svd}. Equality between these vectors, $\vv_{(1),\ell_1}=\vv_{(2),\ell_2}$,  implies a shared joint structure along the selected directions. 
For candidate indices  $\ell_k \in [r_k]$, we consider the following hypotheses: 
\begin{equation}
    \label{eq:two-block-hyp}
H_0: \vv_{(1),\ell_1}=\vv_{(2),\ell_2}
\qquad\textnormal{vs.}\qquad
H_1: \vv_{(1),\ell_1}\neq \vv_{(2),\ell_2}.
\end{equation}
Throughout, the sign ambiguity of the singular vectors is ignored, i.e. we use the convention that $\vv_{(1),\ell_1}=\vv_{(2),\ell_2}$ is understood as $\vv_{(1),\ell_1}= \pm\vv_{(2),\ell_2}$.  

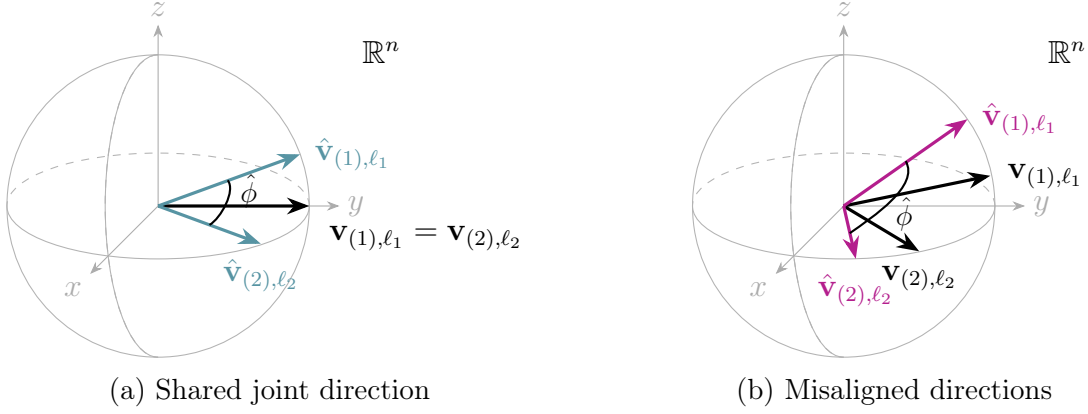
\begin{figure}[t]
\centering
\begin{subfigure}[t]{0.47\textwidth}
\centering
\begin{tikzpicture}[scale=2.0, >=Stealth]

  % Sphere outline
  \draw[thin,gray!60] (0,0) circle (1);
  \def\e{0.35}
  \draw[thin,gray!60] (-1,0) arc[start angle=180,end angle=360,x radius=1,y radius=\e];
  \draw[thin,gray!60,dashed] (1,0) arc[start angle=0,end angle=180,x radius=1,y radius=\e];
  \draw[thin,gray!60] (0,-1) arc[start angle=270,end angle=90,x radius=\e,y radius=1];
  \draw[thin,gray!60,dashed] (0,1) arc[start angle=90,end angle=270,x radius=\e,y radius=1];

  % Axes
  \draw[->,gray!65] (0,0)--(-0.45,-0.45) node[below left=-1pt]{$x$};
  \draw[->,gray!65] (0,0)--(1.20,0) node[right=-1pt]{$y$};
  \draw[->,gray!65] (0,0)--(0,1.20) node[above=-1pt]{$z$};

  % True joint direction v_J
  \draw[very thick,black,->] (0,0)--(1,0) node[below right=3pt]{$\vv_{(1),\ell_1}=\vv_{(2),\ell_2}$};

  % Noisy estimates (aligned)
  \def\cosa{0.9397}\def\sina{0.3420}
  \draw[very thick,red!17!green!50!blue!72,->] (0,0)--(\cosa,\sina)
    node[right=1pt]{$\hat\vv_{(1),\ell_1}$};
  \draw[very thick,red!17!green!50!blue!72,->] (0,0)--(0.68,-0.25)
    node[below=1pt]{$\hat\vv_{(2),\ell_2}$};

  % Angle phi
  \draw[thick] (0.34,-0.125) to[out=30,in=-50,looseness=0.7](0.46,0.17);
  \node[scale=0.9] at (0.60,0.10){$\hat\phi$};

  \node[anchor=west] at (1.28,1.02) {$\mathbb{R}^n$};

\end{tikzpicture}
\caption*{(a) Shared joint direction}
\end{subfigure}
\hfill
\begin{subfigure}[t]{0.47\textwidth}
\centering
\begin{tikzpicture}[scale=2.0, >=Stealth]

  % Sphere outline
  \draw[thin,gray!60] (0,0) circle (1);
  \def\e{0.35}
  \draw[thin,gray!60] (-1,0) arc[start angle=180,end angle=360,x radius=1,y radius=\e];
  \draw[thin,gray!60,dashed] (1,0) arc[start angle=0,end angle=180,x radius=1,y radius=\e];
  \draw[thin,gray!60] (0,-1) arc[start angle=270,end angle=90,x radius=\e,y radius=1];
  \draw[thin,gray!60,dashed] (0,1) arc[start angle=90,end angle=270,x radius=\e,y radius=1];

  % Axes
  \draw[->,gray!65] (0,0)--(-0.45,-0.45) node[below left=-1pt]{$x$};
  \draw[->,gray!65] (0,0)--(1.20,0) node[right=-1pt]{$y$};
  \draw[->,gray!65] (0,0)--(0,1.20) node[above=-1pt]{$z$};

  % True directions (different)
  \draw[very thick,black,->] (0,0)--(0.97,0.2) node[right=2pt]{$\vv_{(1),\ell_1}$};
  \draw[very thick,black,->] (0,0)--(0.5,-0.3) node[below=1pt]{$\vv_{(2),\ell_2}$};

 % Noisy estimates (aligned)
  \def\cosa{0.819}\def\sina{0.574}
  \draw[very thick,magenta!70!blue,->] (0,0)--(\cosa,\sina)
    node[right=1pt]{$\hat\vv_{(1),\ell_1}$};
  \draw[very thick,magenta!70!blue,->] (0,0)--(0.08,-0.35)
    node[below=1pt]{$\hat\vv_{(2),\ell_2}$};

  % Angle phi
  \draw[thick] (0.04,-0.18) to[out=35,in=-60,looseness=0.8](0.41,0.297);
  \node[scale=0.9] at (0.4,-0.05){$\hat{\phi}$};
  
  \node[anchor=west] at (1.28,1.02) {$\mathbb{R}^n$};
  
\end{tikzpicture}
\caption*{(b) Misaligned directions}
\end{subfigure}

\caption{
Illustration of the empirical alignment angle $\hat\phi = \hat\phi_{\ell_1,\ell_2}$ under the null and the alternative.}
\label{fig:H0vsH1}
\end{figure}

We test these hypotheses using the multi-source data $(\Xv_1, \Xv_2)$, obtained under the integrative multi-source factor model \eqref{model:jive}. Since the hypotheses can also be written in terms of the angle formed by two directions $\phi_{\ell_1, \ell_2}:= \angle\!\left(\vv_{(1),\ell_1},\,\vv_{(2),\ell_2}\right)$, we measure empirical
 alignment by the angle between the sample right-singular vectors:
$$
\hat{\phi}_{\ell_1,\ell_2}
:= \angle\!\left(\hat{\vv}_{(1),\ell_1},\,\hat{\vv}_{(2),\ell_2}\right), \quad \hat{\vv}_{(k),\ell_k} := \vv_{\ell_k}(\Xv_k), 
$$
where $\vv_{i}(\Xv)$ stands for the $i$-th singular vector of $\Xv$. Here, the angle $\angle(\uv,\vv)$ is understood as the acute angle between two axes, i.e. $\angle(\uv,\vv) =\cos^{-1}(|\uv^\top\vv|)$.
Figure~\ref{fig:H0vsH1} illustrates that the empirical alignment angle $\hat{\phi}_{\ell_1,\ell_2}$ is typically smaller under the null hypothesis than under the alternative. We construct a test procedure based on the following theorem, which characterizes the asymptotic distribution of $\cos\hat{\phi}_{\ell_1,\ell_2}$ under $H_0$.

% ================================================================
% ==== Figure: Two noisy estimates ===============================

%====================================================================================

\begin{thm}[Asymptotic null distribution of the empirical alignment]
\label{thm:dist.phi.two}
Suppose Assumption~\ref{assump:high-dim_supercritical} holds for $\Xv_1$ and $\Xv_2$. Under $H_0$, i.e., $\vv_{(1),\ell_1}=\vv_{(2),\ell_2}$, as $n, p_1, p_2 \to \infty$: 
\begin{align}\label{eq:thm.phi.block2}
    \sqrt{\frac{n}{\Vc_{y_1,y_2}(\lambda_{(1),\ell_1},\,\lambda_{(2),\ell_2})}}
\!\left\{
\cos\hat{\phi}_{\ell_1,\ell_2} - m_{y_1}(\lambda_{(1),\ell_1})\,m_{y_2}(\lambda_{(2),\ell_2})
\right\}
\;\xrightarrow{d}\; N(0,1),
\end{align}
where %the $\Vc_{y_1,y_2}(\lambda_1,\lambda_2)$ is defined as follows:
    \begin{align*}
        &m_y(s) := \sqrt{\frac{s^4-y}{s^2(s^2+1)}}, \quad\theta_y(s) :=\frac{s^4+2 y s^2+y}{s^3\left(s^2+1\right)^2}, \quad \psi_y(s):=\frac{s^6-3 y s^2-2 y}{s^3\left(s^2+1\right)^2}, \\
        &\mathcal{V}_y(s):=  \frac{s^2(s^2+1)}{2(s^4-y)^2} \left(2 y(y+1) \theta_y(s)^2-\frac{y(y-1)(5 y+1)}{s\left(s^2+1\right)^2} \theta_y(s)\right. \\
        &\quad\quad\quad\quad\quad\quad\quad\quad\quad\quad\left.+\frac{\left(s^4+y\right)\left(s^2+y\right)^2}{s^3\left(s^2+1\right)^2} \psi_y(s) +\frac{2 y^2(y-1)^2}{s^2\left(s^2+1\right)^4}\right) + \frac{\theta^2_y(s)}{m_y^2(s)} , \\
        &\mathcal{V}_{y_1,y_2}(s_{1},s_{2}) = m_{y_1}^2(s_1)\mathcal{V}_{y_1}(s_{1}) + m_{y_2}^2(s_2)\mathcal{V}_{y_2}(s_{2}) 
    + (1-m^2_{y_1}(s_{1}) )(1-m^2_{y_2}(s_{2}) ).
    \end{align*}
\end{thm}

We note that the expressions $m_{y_k}(\lambda_{(k),\ell_k})$ and $\mathcal{V}_{y_k}(\lambda_{(k),\ell_k})$ are the mean and $n$-scaled variance of the limiting distribution of $|\vv_{(k),\ell_k}^\top\hat\vv_{(k),\ell_k}|$ \citep[][see also Lemma~\xdocref{S4} in Supplementary \xdocref{S2.2}]{bao2021singular}. 
These asymptotic moments depend on the unknown SNR parameters $\lambda_{(1),\ell_1}$ and $\lambda_{(2),\ell_2}$, while the aspect ratios $y_k = p_k/n$ are explicitly known. 
Replacing the unknown parameters by their consistent estimators 
$\hat{\lambda}_{(k),\ell_k}$, defined later in Section~\ref{sec:estimation}, we use the following standardized statistic to test the alignment (\ref{eq:two-block-hyp}): %, we use the standardized statistic:
\[
\widehat{Z}_{\ell_1,\ell_2}
:=\sqrt{\frac{n}{\Vc_{y_1,y_2}(\hat{\lambda}_{(1),\ell_1},\,\hat{\lambda}_{(2),\ell_2})}}
\left\{
\cos\hat{\phi}_{\ell_1,\ell_2}
-{m_{y_1}(\hat{\lambda}_{(1),\ell_1})\,m_{y_2}(\hat{\lambda}_{(2),\ell_2})}
\right\}.
\]
For a significance level $\alpha\in(0,1)$, we reject $H_0$ if $\widehat{Z}_{\ell_1,\ell_2}\le - z_{\alpha}$, where $z_{\alpha}=\Phi^{-1}(1-\alpha)$. 

\begin{figure}[t]
    \centering
    \includegraphics[width=0.8\linewidth]{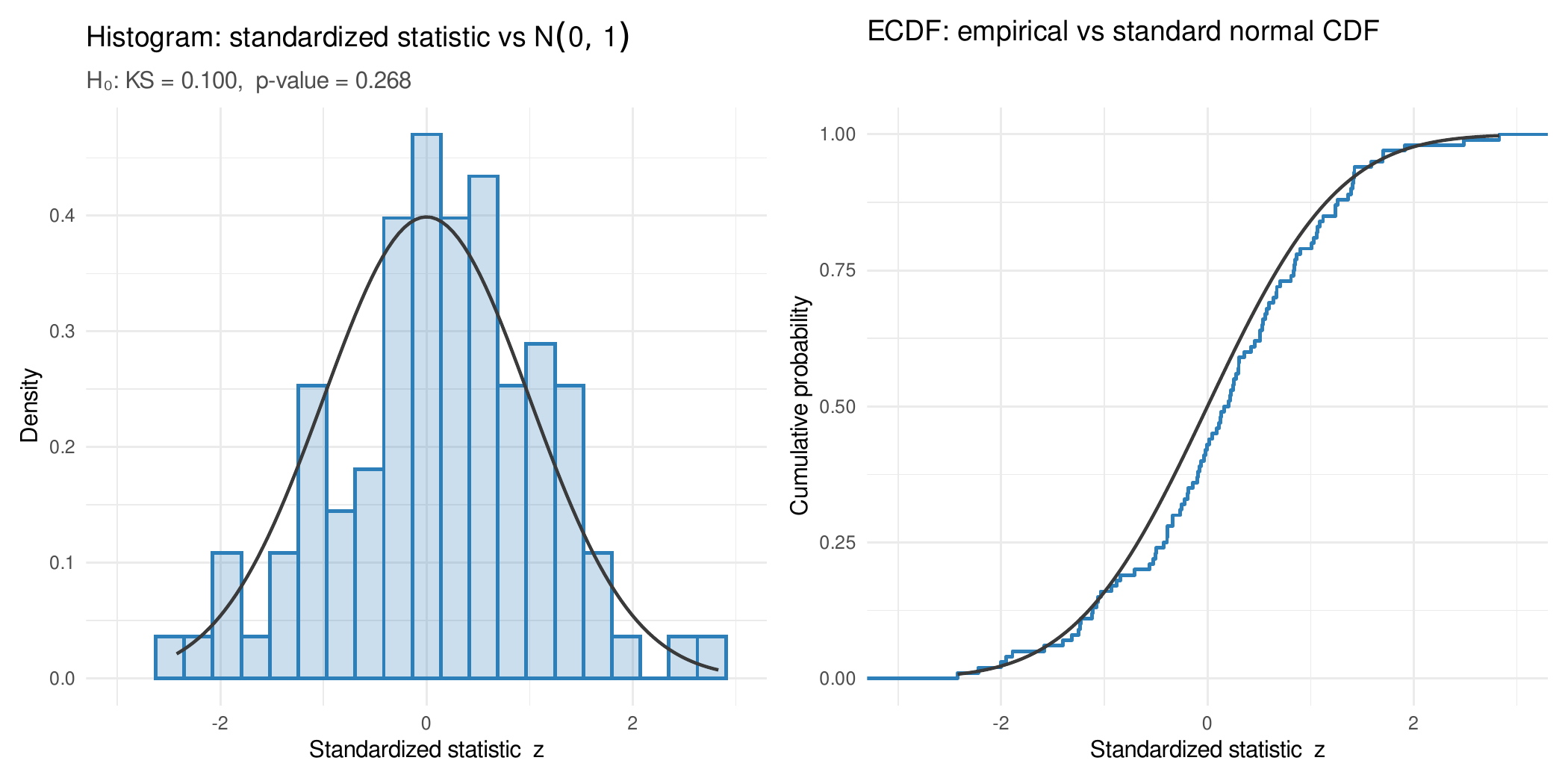}
    \caption{Finite-sample null distribution of the statistic $\widehat{Z}_{\ell_1,\ell_2}$ is well approximated by the standard normal distribution.}
    \label{fig:asym.dist.phi.block2}
\end{figure}

To evaluate the empirical accuracy of the normal approximation, we conducted numerical simulations with $(n,p_1,p_2) = (1000, 800, 600)$. Figure \ref{fig:asym.dist.phi.block2} presents the results from $N=100$ independent trials, showing that the empirical distribution of the standardized statistic $\widehat{Z}_{\ell_1,\ell_2}$ aligns well with the standard normal distribution. In particular, a Kolmogorov--Smirnov test failed to reject the null hypothesis of normality at the conventional significance level. See Supplementary~\xdocref{S1.1} for a detailed parameter setting. 

\subsection{Extensions to multi-block cases}\label{subsubsec:multi_block}

We now generalize the two-block alignment test to an arbitrary number of data blocks $K > 2$. Suppose we have a set of candidate right-singular vectors $\{\vv_{(k),\ell_k}\}_{k=1}^K$, which are identified as potential constituents of a shared joint component.  To evaluate their collective alignment across all blocks, we consider the following hypotheses: 
 \begin{equation}
    \label{eq:hyp_multi}
H_0: \vv_{(1),\ell_1}=\vv_{(2),\ell_2} = \cdots  = \vv_{(K),\ell_K}
\quad\textnormal{vs.}\quad
H_1: \vv_{(k),\ell_k}\neq \vv_{(l),\ell_l} \,\,\mbox{for some} \, \,k\neq l.
\end{equation}

To quantify joint alignment across the $K$ estimated directions $\{\hat{\vv}_{(1),\ell_1}, \dots, \hat{\vv}_{(K),\ell_K} \}$, we define a joint alignment statistic as the squared norm of their sum:
\begin{align*}
    T^2_{\boldsymbol\ell}
    := \big\| \hat{\vv}_{(1),\ell_1} + \hat{\vv}_{(2),\ell_2} + \cdots + \hat{\vv}_{(K),\ell_K} \big\|_2^2,
    \quad \boldsymbol\ell=\{\ell_1,\dots,\ell_K\}.
\end{align*}
In practice, since the sign of each $\hat{\vv}_{(k),\ell_k}$ is arbitrary, we maximize the statistic over all possible sign configurations: $T^2_{\boldsymbol\ell} = \max_{\epsilon_k \in \{1,-1\}} \|\sum_{k=1}^K \epsilon_k \hat{\vv}_{(k),\ell_k}\|^2$. Under the null hypothesis, all block-specific estimates cluster around the common direction $\vv_\textsc{j} := \vv_{(k),\ell_k}$ (for all $k$), leading to a large value of $T^2_{\boldsymbol\ell}$. 
Conversely, when the directions are dispersed across blocks, the statistic tends to have smaller values. %he vector sum undergoes partial cancellation, resulting in a smaller value of the statistic. 
This motivates a test procedure that rejects $H_0$ when $T^2_{\boldsymbol\ell}$ falls below a specified threshold.

For the theoretical derivation of the null distribution, we fix the signs such that 
\begin{equation}
    \langle \vv_{\textsc{j}}, \hat{\vv}_{(k),\ell_k}\rangle \ge 0, \quad k \in [K].
    \label{eq:sign_convention}
\end{equation}
 This convention is justified by the fact that $\langle \vv_{\textsc{j}}, \hat{\vv}_{(k),\ell_k}\rangle^2 = m^2_{y_k}(\lambda_{(k),\ell_k}) + O_p(n^{-1/2})$ as $n, p_k \to \infty$ under $H_0$ \citep[][see also Lemma~\xdocref{S4} in Supplementary \xdocref{S2.2}]{bao2021singular}. For large sample sizes, the sign configuration that maximizes $T^2_{\boldsymbol\ell}$ effectively aligns all empirical directions toward the common vector $\vv_\textsc{j}$, satisfying the positive inner product convention (\ref{eq:sign_convention}). 
The following theorem characterizes the asymptotic distribution of $T^2_{\boldsymbol{\ell}}$ under $H_0$.

\begin{thm}[Asymptotic null distribution of the joint  alignment]\label{thm:dist.phi.multi}

Suppose Assumption~\ref{assump:high-dim_supercritical} holds for each $\Xv_k$, $k \in [K]$, and the signs of empirical singular vectors are chosen to satisfy (\ref{eq:sign_convention}). Under $H_0$, as $n, p_k \to \infty$ for all $k$:
\begin{align*}
    Z_{\boldsymbol\ell} := \sqrt{\frac{n}{\Vc_{\yv}^\circ(\lambdav)}}
    \Big\{T^2_{\boldsymbol\ell}-\mu_{\yv}(\lambdav)\Big\}
    \;\xrightarrow{d}\; N(0,1),
\end{align*}
where 
$\mu_{\yv}(\lambdav)
= K + 2\sum_{k< l}{m_{y_k}(\lambda_{(k),\ell_k})\,m_{y_l}(\lambda_{(l),\ell_l})}$,
and
\[
\Vc_{\yv}^\circ(\lambdav)
= 4\sum_{k=1}^K {\Vc_{y_k}(\lambda_{(k),\ell_k})}
\left\{\sum_{l\neq k}{m_{y_l}(\lambda_{(l),\ell_l})}\right\}^2
+ 4\sum_{k< l}\big(1-m^2_{y_k}(\lambda_{(k),\ell_k})\big)\big(1-m^2_{y_l}(\lambda_{(l),\ell_l})\big).
\]
\end{thm}

Theorem~\ref{thm:dist.phi.multi} generalizes  Theorem~\ref{thm:dist.phi.two}, to which it reduces in the specific case of $K=2$. The joint alignment statistic  aggregates all pairwise alignment information, which can be seen from the alternative expression of the statistic: $T^2_{\boldsymbol\ell} = K + 2\sum_{k < l} \cos\hat{\phi}_{\ell_k,\ell_l}$. Furthermore, the limiting mean $\mu_{\yv}(\lambdav)$ exhibits a similar structure where each inner product $\cos\hat{\phi}_{\ell_k,\ell_l}$ is replaced by its respective asymptotic limit.

To test the joint alignment (\ref{eq:hyp_multi}), we use a standardized statistic with the unknown SNR parameters ${\lambda}_{(k),\ell_k}$ replaced by their consistent estimators $\hat{\lambda}_{(k),\ell_k}$: 
\begin{equation}
    \label{eq:std.stat.multi}
    \widehat{Z}_{\boldsymbol\ell}
= \sqrt{\frac{n}{{\Vc}^\circ_{\yv}(\widehat{\lambdav})}}
\left\{T^2_{\boldsymbol\ell}-\mu_{\yv}(\widehat{\lambdav})\right\}.
\end{equation}
The null hypothesis is  rejected at significance level $\alpha$ if $\widehat{Z}_{\boldsymbol\ell}\le - z_{\alpha}$. 

\begin{figure}[t]
    \centering
    \includegraphics[width=0.9\linewidth]{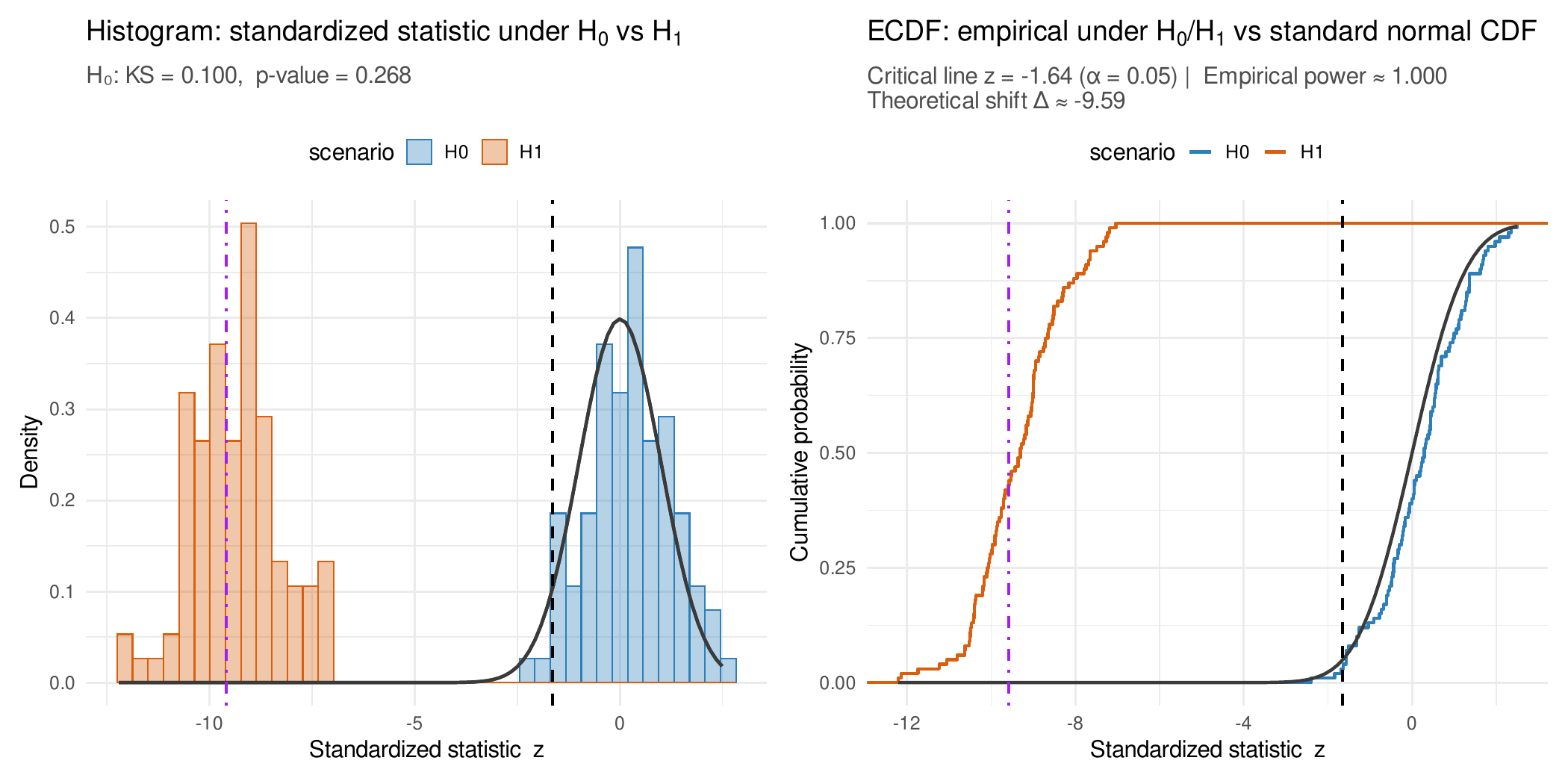}
    \caption{
    Empirical distributions of $\widehat{Z}_{\boldsymbol{\ell}}$ \eqref{eq:std.stat.multi} under $H_0$ and $H_1$. The vertical dashed line marks the threshold $-z_{0.05}$, and the dot-dashed line indicates the shift $\Delta_{\boldsymbol\ell}^{\textnormal{st}}$ under $H_1$.}
    \label{fig:asym.dist.phi.multi}
\end{figure}

Theoretically, if the estimators satisfy $\hat{\lambda}_{(k),\ell_k} = {\lambda}_{(k),\ell_k} + o_p(n^{-1/2})$, the proposed test maintains an asymptotic size of $\alpha$. However, the estimator defined later in Section~\ref{sec:estimation} is only shown to satisfy $\hat{\lambda}_{(k),\ell_k} = \lambda_{(k),\ell_k} + O_\prec(n^{-1/2})$ (See Lemma~\ref{lem:estimation}). Our numerical results suggest that the difference $\sqrt{n}(\lambda_{(k),\ell_k} - \hat{\lambda}_{(k),\ell_k})$ is typically small but positive; consequently, the empirical size turns out to be slightly smaller than $\alpha$. Thus, while the test is somewhat conservative, it effectively controls the type I error; see Figure~\ref{fig:asym.dist.phi.multi}.

On the other hand, the test has asymptotic power one against $n^{-1/2}$-separated alternatives. 

\begin{thm}[Asymptotic power one]
\label{thm:power_one}
    Suppose Assumption~\ref{assump:high-dim_supercritical} holds for each $\Xv_k$, $k \in [K]$. 
    Suppose that there exists a pair $\{k,l\} \subset [K]$ such that, as $n,p_k\to\infty$,  
$\sqrt{n} (1-\cos{\phi_{\ell_k,\ell_l}}) = \sqrt{n}(1-\vv_{(k),\ell_k}^\top \vv_{(l),\ell_l}) \to \infty$. Then for any $\alpha \in (0,1)$, 
$$
\lim_{n,p_k\to\infty} \mathbb{P}( \widehat{Z}_{\boldsymbol\ell} \le -z_\alpha) = 1.
$$
\end{thm}
 
To understand this result, we characterize the shift of the statistic under $H_1$. The 
% limiting mean 
asymptotic centering of the statistic $T^2_{\boldsymbol\ell}$ is 
$\mu_{\yv}(\widehat{\lambdav},\boldsymbol\ell) - \Delta_{\boldsymbol\ell}$, where  
\[
\Delta_{\boldsymbol\ell} = 2\sum_{k<l}{m_{y_k}(\lambda_{(k),\ell_k})\,m_{y_l}(\lambda_{(l),\ell_l})}\,\bigl(1-\cos\phi_{\ell_k,\ell_l}\bigr).
\]
See Lemma \xdocref{S5} in Supplementary~\xdocref{S2.3}.
Since $\Delta_{\boldsymbol\ell} > 0$ under $H_1$, the standardized statistic $\widehat{Z}_{\boldsymbol\ell}$ is expected to shift to the left by an amount $\Delta_{\boldsymbol\ell}^{st} = \Delta_{\boldsymbol\ell}\sqrt{n/{{\Vc}^\circ_{\yv}({\lambdav})}}$. An example of the shift under $H_1$ for a three-block data situation is illustrated in  Figure~\ref{fig:asym.dist.phi.multi}; see Supplementary~\xdocref{S1.1} for a detailed parameter setting for the example. 

\section{Method}\label{sec:method}
This section describes the proposed Multi-Source Sequential Alignment Test (MSSAT) for decomposing multi-source data into joint and individual components. The proposed framework comprises three  stages: (i) estimation of signal ranks and noise levels, (ii) determination of the joint rank via sequential alignment testing, and (iii) reconstruction of the signal matrices.  We also extend this methodology to identify partially shared components across subsets of data sources.

\subsection{Signal rank and noise level estimation}\label{sec:estimation}

To conduct inter-block alignment inference, we first estimate, for each block $k$, the signal rank $r_k$, the noise variance $\sigma_k^2$, the signal strengths $\{d_{(k),l}\}_{l=1}^{r_k}$, and the SNRs $\lambda_{(k),l}=d_{(k),l}/(\sqrt{n}\sigma_k)$.

For estimation of the noise variance and the signal rank, we adopt the bulk eigenvalue matching (BEMA) method of \citet{ke2023estimation}. Once the rank and noise levels are determined, we employ the standard de-biasing method \citep{benaych2012singular} to estimate individual signal strength $d_{(k),l}$'s. 

The choice of these estimators is rooted in random matrix theory. It is well known that under the spiked covariance model, the empirical distribution of the non-spike (bulk) eigenvalues of the sample covariance matrix converges to the Marchenko--Pastur (MP) law, while the largest bulk eigenvalue exhibits Tracy--Widom (TW) fluctuations at the upper edge \citep[see][and references therein]{bloemendal2016principal}. A closely analogous phenomenon holds under our additive noise model \eqref{model:jive}: the empirical distribution of the squared singular values associated with the noise component similarly converges to the MP law \citep{bai2010spectral}, and its upper edge obeys TW fluctuations \citep{ding2022tracy}. These similarities justify adapting the BEMA framework to our setting. While the consistency of BEMA estimators has been established exclusively under the spiked covariance model \citep{ke2023estimation}, we establish that the same estimators remain consistent under our noise-additive model \eqref{model:jive}. These theoretical results are formally presented in Lemma~\ref{lem:estimation} following the definition of the estimators below.

 \paragraph{Estimation of noise variance via bulk spectrum fitting.}
Let $\sigma_1(\Xv_k)\ge\cdots\ge\sigma_{m_k}(\Xv_k)$ denote the singular values of $\Xv_k$, where $m_k=\min(p_k,n)$.
For a fixed trimming parameter $\beta\in(0,1/2)$, 
define the index range $ (m_{k,\mathrm{low}}(\beta),m_{k,\mathrm{up}}(\beta))= (\beta m_k, (1-\beta) m_k)$. %$ and $m_{k,\mathrm{up}}(\beta) = (1-\beta) m_k$.
Let $q_{k,l}$ be the $(l/m_k)$ upper-quantile of $\operatorname{MP}(y_k, 1)$, the Marchenko--Pastur distribution with aspect ratio $y_k$ and unit scale. We estimate $\sigma_k^2$ by
\begin{align}\label{equation:est.noise}
    \hat{\sigma}_k^2 = \frac{\sum_{m_{k,\mathrm{low}}(\beta) \leq l \leq m_{k,\mathrm{up}}(\beta)}q_{k,l} \sigma_l^2(\Xv_k)}{\sum_{m_{k,\mathrm{low}}(\beta) \leq l \leq m_{k,\mathrm{up}}(\beta)}q_{k,l}^2}.
\end{align}
This estimator fits the MP law by matching a central $(1-2\beta)$ fraction of the squared empirical singular values,
$\{\sigma_l^2(\Xv_k):\, l\in[m_{k,\mathrm{low}}(\beta),\,m_{k,\mathrm{up}}(\beta)]\}$. This approach of aggregating mid-spectrum bulk outperforms strategies that match a single point, e.g. the median-matching method of \citet{gavish2014optimal}. Throughout our numerical experiments, we fix $\beta = 0.2$.

\paragraph{Estimation of signal rank.}
Fix a tuning parameter $\gamma \in (0,1)$ controlling the degree of overestimation of $r_k$, and let $\operatorname{TW}_{1-\gamma}$ denote the $(1-\gamma)$-quantile of the Tracy--Widom distribution.
We estimate $r_k$ by counting the number of squared singular values that exceed the calibrated bulk edge:
\begin{align}\label{equation:est.rank}
    \hat{r}_k = \# \left\{ l \in [m_k]: \sigma_l^2(\Xv_k)/n > \hat{\sigma}_k^2\left[(1+\sqrt{y_k})^2 + \operatorname{TW}_{1-\gamma}n^{-\frac{2}{3} }y_k^{-\frac{1}{6}} (1+\sqrt{y_k})^{\frac{4}{3}}   \right] \right\}.
\end{align}
We use $\gamma = 0.1$ in our numerical experiments.

\paragraph{Estimation of signal singular values and SNRs.}
After determining $(\hat{r}_k,\hat{\sigma}_k^2)$, we estimate the population singular values of the signal matrix. Since the empirical singular values $\sigma_l(\Xv_k)$ are upward-biased, we invert the deterministic limit mapping between population and sample singular values \citep{benaych2012singular}.
Specifically, for each $l \in [\hat{r}_k]$, the debiased estimator of the signal singular value is 
\begin{align}\label{equation:benaych}
\hat{d}_{(k),l}^2 = \frac{n\hat{\sigma}_k^2}{2}\left(  \frac{\sigma_l^2\left( \Xv_k \right)}{n\hat{\sigma}_k^2} -1-y_k + \sqrt{\left(\frac{\sigma_l^2\left( \Xv_k \right)}{n\hat{\sigma}_k^2} -1-y_k\right)^2 -4y_k}\right).
\end{align}
The corresponding signal-to-noise ratio (SNR) is estimated by 
\begin{align}\label{equation:est.SNR}
    \hat{\lambda}_{(k),l} = \frac{\hat{d}_{(k),l}}{\sqrt{n}\hat{\sigma}_k}.
    % \hat{\lambda}_{(k),l}=\hat{d}_{\Av_k,l}/(\sqrt{n}\hat{\sigma}_k).
\end{align}

The following lemma establishes the consistency of our estimators \eqref{equation:est.noise}, \eqref{equation:est.rank}, and \eqref{equation:est.SNR} under the high-dimensional noise-additive models.

\begin{lem}\label{lem:estimation}
    Under Assumption~\ref{assump:high-dim_supercritical}, if $\beta\in (0,1/2)$ is constant and $\gamma_n \to 0$ at a suitable rate such that $n^{-2/3}\operatorname{TW}_{1-\gamma_n} \to 0$ and $n^{-\epsilon}\operatorname{TW}_{1-\gamma_n} \to \infty$ for some $\epsilon>0$, then we have
    \begin{enumerate}
        \item [(i)] $|\hat{\sigma}_k^2 - \sigma_k^2| = O_\prec(n^{-1})$ for all $k \in [K]$.
        \item [(ii)] $\Pb (\hat{r}_k = r_k) \to 1$ as $n\to \infty$.
        \item [(iii)] $|\hat{\lambda}_{(k),l} - {\lambda}_{(k),l}| = O_\prec(n^{-1/2})$  for all $l \in [r_k]$ and $k \in [K]$.
    \end{enumerate}
\end{lem}

Here, the notation $O_\prec$ denotes the relation of \textit{stochastic domination} introduced by \citet{erdHos2013averaging}.

\subsection{Joint rank estimation via sequential alignment tests}\label{subsec:seq_align_test}

We now address the identification of the shared joint components $\Jv_k$ 
%in \eqref{model:jive}
and the estimation of their common rank $r_{\textsc{j}}$.
 These components share a row subspace $\Jc$ defined by the intersection of the individual signal row subspaces: 
$$
\Jc := \bigcap_{k=1}^K \operatorname{span}(\Vv_{(k)}) \equiv  \text{row}(\Jv_k) ,\, \, \mbox{for all}\,\, k \in [K]. 
$$ 

To estimate $r_{\textsc{j}}$ from the noisy signal subspaces $\hat\Vv_{(k)}$, we sequentially test whether a set of source-specific candidate directions 
$\{\tilde\vv_{(k)} \in \operatorname{span}(\hat\Vv_{(k)}): k \in [K]\}$ represents an empirical realization of a shared joint direction $\vv_{\textsc{j}}$.

The candidate directions are derived by relaxing the assumption of exact alignment with block-wise singular vectors.
If $\mathbf{v}_{\textsc{j}}$ coincides with specific block-wise singular vectors (i.e. for all $k \in [K]$, $\vv_{(k),\ell_k} = \vv_{\textsc{j}}$ for some $\ell_k$), the alignment test from Section~\ref{sec:theory} can be directly applied.
However, $\vv_{\textsc{j}}$ generally lies within the intersection of subspaces without necessarily aligning with any single singular vectors, and we write $\vv_{\textsc{j}} = \Vv_{(k)}\wv_{(k)}$ for some weight vector satisfying $\|\wv_{(k)}\|_2 = 1$. If these weights $\wv_{(k)}$'s were known, an oracle candidate would be
\begin{equation}
    \label{eq:oracle_v}
    \tilde{\vv}_{(k)}^* = \hat\Vv_{(k)}\wv_{(k)}^\circ, \quad \wv_{(k)}^\circ = \Dv_k^{-1}\wv_{(k)} / \|\Dv_k^{-1}\wv_{(k)}\|_2,
\end{equation}
where 
$\Dv_k = \operatorname{diag}(m_{y_k}(\lambda_{(k),1}), \dots, m_{y_k}(\lambda_{(k),r_k}))$ is used to compensate the asymptotic shrinkage of $\hat\Vv_{(k)}^\top \Vv_{(k)}  \overset{p}{\to} \Dv_k$; see Lemma~\xdocref{S4} in the supplementary material.
 The modified weight vector $\wv_{(k)}^\circ$ is scaled to have length 1.

In practice, neither the hypothesized joint direction $\vv_{\textsc{j}}$ nor the weights $\wv_{(k)}$ are known. As a proxy for the joint direction, we identify ``average'' directions of the signal subspaces. Specifically, 
let $\hat\vv_{\textsc{j}}$ be the leading eigenvector of the sum of the projections: 
\begin{equation}
    \label{eq:sum_of_projections}
    \hat{\vv}_{\textsc{j}} = \phi_1(\bar{\Pv}), \quad  \bar{\Pv} = \sum_{k=1}^K \hat{\Pv}_{(k)} = \sum_{k=1}^K \hat\Vv_{(k)}\hat\Vv_{(k)}^\top
\end{equation}
here $\phi_i(\Av)$ stands for the $i$th largest eigenvector of $\Av$. 
While $\hat\vv_{\textsc{j}}$ plays the role of an empirical joint direction (whose counterpart is $\vv_{\textsc{j}}$), we set the empirical weight vector as $\hat\wv_{(k)} = \hat\Vv_{(k)}^\top\hat{\vv}_{\textsc{j}} / \|\hat\Vv_{(k)}^\top\hat{\vv}_{\textsc{j}}\|$. The source-specific candidate directions are then given by replacing the oracle weight  in (\ref{eq:oracle_v}) with the empirical weight: 
\begin{equation}
\label{eq:candidate_v_tilde}
   \tilde{\vv}_{(k)} =\hat\Vv_{(k)}\hat{\wv}^\circ_{(k)},  \quad k \in [K],
\end{equation} 
where
$\hat\wv_{(k)}^\circ = \hat\Dv_k^{-1}\hat\wv_{(k)} / \|\hat\Dv_k^{-1}\hat\wv_{(k)}\|$, and $\hat\Dv_k = \operatorname{diag}(m_{y_k}(\hat\lambda_{(k),1}), \dots, m_{y_k}(\hat\lambda_{(k),r_k}))$.

This choice of candidate directions is motivated by the fact that the matrix ${\Pv} = \sum_{k=1}^K \Vv_{(k)}\Vv_{(k)}^\top$ possesses exactly $r_{\textsc{j}}$ largest eigenvalues equal to $K$, with its $r_{\textsc{j}}$-dimensional eigenspace coinciding with $\mathcal{J}$. %Furthermore, 

To assess the alignment of the candidate set $\{\tilde\vv_{(k)} \in \operatorname{span}(\hat\Vv_{(k)}): k \in [K]\}$ we employ the standardized statistic:
\begin{equation}
    \label{eq:seqtest_procedure}
        \tilde{Z}
= \sqrt{\frac{n}{{\tilde\Vc}^\circ_{\yv}(\widehat{\lambdav})}}
\left\{T^2-\tilde\mu_{\yv}(\widehat{\lambdav})\right\}, \quad T^2 = \| \sum_{k=1}^K \tilde\vv_{(k)}\|^2.
\end{equation} 
The limiting mean $\tilde\mu_{\yv}(\widehat{\lambdav})$ and variance ${\tilde\Vc}^\circ_{\yv}(\widehat{\lambdav})$ of the statistic $T^2$  will be discussed shortly. 
The decision rule is defined as follows:  For a prespecified threshold $t<0$, we reject the null hypothesis of a shared joint direction if $\tilde{Z} < t$, in which case we conclude that there is \textit{no} joint component. Conversely, if $\tilde{Z} \ge t$, we conclude that \textit{at least} one joint component exists. 

The alignment test is applied sequentially to the candidate joint directions derived from the subsequent eigenvectors of $\bar{\Pv}$. 
Specifically, for $i = 2, 3, \dots, \min\{\hat{r}_k\}$, we set $\hat\vv_{\textsc{j}}$ as $\hat{\mathbf{v}}_{{\textsc{j}},i} = \phi_i(\bar{\mathbf{P}})$, the $i$th eigenvector of $\bar{\mathbf{P}}$, and repeat the procedure until the null hypothesis of a shared direction is first rejected. 

The estimated joint rank $\tilde{r}_{\textsc{j}}$ is then the number of acceptances of the sequential alignment tests, and the joint score matrix is estimated as $\tilde\Vv_{\textsc{j}} = [\tilde{\vv}_{{\textsc{j}},1},\ldots, \tilde{\vv}_{{\textsc{j}},\tilde{r}_{\textsc{j}}}]$. The procedure of joint rank estimation is summarized in Algorithm~\ref{alg:sequential}.

\setlength{\algomargin}{1.5em}
\begin{algorithm}[t]
\DontPrintSemicolon
\KwInput{Data blocks $\Xv_k\in\Rb^{p_k\times n}$, signal ranks $r_k$, signal-to-noise ratios $\lambda_{(k),\ell}$ for $\ell \in [r_k]$ and $k \in [K]$, and a threshold $t$.}
\KwOutput{Estimated joint basis $\tilde{\Vv}_{\Jv}$ and joint rank $\tilde r_{\textsc{j}}$.}

Set $\tilde{r} = \min_{k\in[K]}\{r_k\}$ and compute $\hat{\Vv}_{(k)}$, the rank-$r_k$ right singular matrix of each $\Xv_k$, and form $\bar{\Pv} = \sum_{k=1}^K \hat{\Pv}_{(k)}$.\\

\For{$s=1,\dots,\tilde r$}{
    Set $\tilde{\vv}_{\textsc{j},s}$ to the $s$-th eigenvector of $\bar{\Pv}$.
    
    \ForEach{$k=1,\dots,K$}{
    Compute the weights
    $\hat{\wv}_{k}$ and $\hat{\wv}_{k}^\circ$,
    and mean and variance
    $\tilde \mu_{\yv}(\lambdav)$,
    $\tilde \Vc^\circ_\yv(\lambdav)$.\\
    Compute the source-specific candidate directions $\tilde{\vv}_{(k)}$.}
    
    Evaluate the estimated standardized statistic $\tilde{Z}$. \\
    \If{$ \tilde{Z} < t$}{
    \textbf{Stop.} Set $\tilde r_{\textsc{j}} = s-1$ and break the loop.}
}

If no rejection occurs up to $s=\tilde r$, set $\tilde r_{\textsc{j}}=\tilde r$.\\
Collect all accepted directions
$\{\tilde{\vv}_{{\textsc{j}},1},\dots,\tilde{\vv}_{{\textsc{j}},\tilde r_{\textsc{j}}}\}$
to construct the estimated joint basis
$\tilde{\Vv}_{\Jv} = [\,\tilde{\vv}_{{\textsc{j}},1},\dots,\tilde{\vv}_{{\textsc{j}},\tilde r_{\textsc{j}}}\,]$.

\caption{Joint component estimation via sequential alignment tests.}
\label{alg:sequential}
\end{algorithm}
\vspace{0.5em}
% % ##################################################################

The limiting mean and variance of $T^2 = \| \sum_{k=1}^K \tilde\vv_{(k)}\|^2$ in (\ref{eq:seqtest_procedure}) are estimated by 
\begin{equation}
\begin{aligned}
\tilde\mu_{\yv}(\widehat{\lambdav}) & = K + 2 \sum_{k < l} \hat{m}_{(k)} \hat{m}_{(l)},  \\
    {\tilde\Vc}^\circ_{\yv}(\widehat{\lambdav}) & = 
    4 \sum_{k=1}^K \hat\Vc_{(k)} (\sum_{l \neq k} \hat{m}_{(k)} )^2 + 4 \sum_{k < l} (1 - \hat{m}_{(k)}^2) ( 1- \hat{m}_{(l)}^2),
\end{aligned}
\label{eq:working_mean_var}
\end{equation}     
where the block-specific empirical moments are given by 
\begin{equation}
\label{eq:k_working_mean_var}
    \hat{m}_{(k)} = \sum_{i=1}^{r_k} \hat{w}_{(k),i} \hat{w}^\circ_{(k),i} m_{y_k}(\hat{\lambda}_{(k),i}),\quad \hat\Vc_{(k)} = \sum_{i=1}^{r_k} (\hat{w}_{(k),i} \hat{w}^\circ_{(k),i})^2 \Vc_{y_k}(\hat{\lambda}_{(k),i}).
\end{equation}
In these expressions, 
$\hat{w}_{(k),i}$ and $\hat{w}_{(k),i}^\circ$ denote the $i$th elements of the empirical weight vectors $\hat\wv_{(k)}$ and $\hat\wv_{(k)}^\circ$, respectively.

Direct evaluation of the limiting distribution of $T^2$ is complicated by the stochastic dependency of the empirical weights. To facilitate the derivation, we utilize the oracle candidates defined in \eqref{eq:oracle_v} as analytical surrogates for their empirical counterparts. Accordingly, the asymptotic results established in the following theorem provide the theoretical motivation for the specific functional forms of the mean and variance adopted in \eqref{eq:working_mean_var}.

\begin{thm}\label{thm:oracle}
Suppose Assumption~\ref{assump:high-dim_supercritical} holds for each $\Xv_k$, and let $\wv_{(k)}$ be given to satisfy $\|\wv_{(k)}\|_2 = 1$ for all $k \in [K]$. For the weighted singular vectors $\vv_{(k),\textsc{j}} = \Vv_{(k)} \wv_{(k)}$ and the oracle source-specific directions $\tilde{\vv}_{(k)}^\ast$ (\ref{eq:oracle_v}), assume that there exist constants $m_{(k)}$ and $\Vc_{(k)}$ such that as $n,p_k \to \infty$, 
$\sqrt{n}( \vv_{(k),\textsc{j}}^\top \tilde{\vv}_{(k)}^\ast  - m_{(k)}) \overset{d}{\to} N(0,\Vc_{(k)}),$
for all $k$. Then the following holds. 
\begin{itemize}
    \item[(a)] As $n,p_k \to \infty$, $T^2_\ast = \|\sum_{k=1}^K \tilde{\vv}_{(k)}^\ast\|^2_2 = m^\circ -\Delta^\circ(\phi) + O_p(n^{-1/2})$, where 
    \begin{equation}
    m^\circ = K + 2 \sum_{k < l} m_{(k)} m_{(l)}, \quad 
        \Delta^\circ(\phi) = 2\sum_{k<l} m_{(k)} m_{(l)} ( 1- \cos\phi_{k,l}),        
    \end{equation} 
    with $\phi_{k,l} = \angle (\Vv_{(k)}\wv_{(k)},\Vv_{(l)}\wv_{(l)})$.
    \item[(b)] If for some $\vv_{\textsc{j}} \in \Jc$, $\vv_{\textsc{j}} = \Vv_{(1)}\wv_{(1)} = \cdots = \Vv_{(K)}\wv_{(K)}$, then as $n,p_k \to \infty$
     \begin{align*}
        \sqrt{\frac{n}{\Vc^\circ}}\left\{  T_\ast^2 - m^\circ \right\} \overset{d}{\to} N(0,1),
    \end{align*}
    where $\Vc^\circ = 4\sum_{k=1}^K \Vc_{(k)}\left\{ \sum_{l\neq k} m_{(l)} \right\}^2 + 4 \sum_{k < l} (1-m^2_{(k)})(1-m^2_{(l)})$.
\end{itemize} 
\end{thm} 

Theorem \ref{thm:oracle}(b) characterizes the limiting distribution of the oracle alignment statistic $T^2_\ast$  under the null hypothesis of global alignment. The resulting functional forms for the mean $m^\circ$ and variance $\Vc^\circ$ serve as the basis for our proposed estimators in \eqref{eq:working_mean_var}.

To justify our estimator of the block-specific moments (\ref{eq:k_working_mean_var}), we utilize the weight representation to expand the inner product:   
\begin{equation}
\vv_{\textsc{j}}^\top \tilde\vv_{(k)}^\ast = \sum_{i=1}^{r_k} w_{(k)_i} w_{(k),i}^\circ\vv_{(k),i}^\top \hat\vv_{(k),i} +  \sum_{i\neq j} w_{(k),i} {w}_{(k),j}^\circ\vv_{(k),i}^\top \hat\vv_{(k),j}.
\label{eq:inner}
\end{equation}

The cross-product terms turn out to be negligible.  
By leveraging the results on the convergence of $\vv_{(k),i}^\top \hat\vv_{(k),j}$, the limiting mean $m_{(k)}$ of $\mathbf{v}_{\textsc{j}}^\top \tilde{\mathbf{v}}_{(k)}^\ast$ coincides with the form of $\hat{m}_{(k)}$ in \eqref{eq:k_working_mean_var} when empirical estimates are replaced by population parameters.
While the exact limiting variance $\mathcal{V}_{(k)}$ is difficult to evaluate, the first term of (\ref{eq:inner}) motivates the expression \eqref{eq:k_working_mean_var}, which provides a good working approximate.  For the purpose of joint rank estimation, the precise determination of the asymptotic mean is significantly more critical  than the exact characterization of the higher-order variance terms. 
Detailed derivations and further discussion on these approximations are provided in Section~\xdocref{S2.5} of the Supplementary Material.

\subsection{Choice of the threshold}\label{sec:threshold}
Selecting an appropriate threshold for $\tilde{Z}$ \eqref{eq:seqtest_procedure} is critical, as the sequential nature of our procedure inherently inflates the Type I error rate. Using a conventional threshold like $-z_{0.05}$ at each step would thus lead to an underestimation of the joint rank. 

Controlling the family-wise Type I error rate (FWER) over a potentially indefinite sequence of tests can be considered as well. One might adopt an alpha spending approach \citep{foster2008alpha,tian2021online}, such as setting the significance level for the $s$th test as $\alpha_s = 2^{-s}\alpha$. While this ensures the overall FWER bounded by $\alpha$, it still permits a 2.5\% probability (at $s=1$ with $\alpha=0.05$) of falsely rejecting a valid shared joint direction at the very first step. Such an early termination still leads to serious underestimation of the joint rank. 
This motivates a more conservative thresholding strategy.

We adopt a $6\sigma$ rule by setting the threshold $t = -6$, which corresponds to a nominal one-sided Type I error rate of approximately $10^{-10}$. 
While such an extremely conservative choice might appear to compromise the power of the test (i.e., by failing to reject $H_0$ in the absence of joint structure), the asymptotic consistency of our test ensures that power approaches one for any $n^{-1/2}$-separated alternative. Consequently, this $6\sigma$ rule performs remarkably well in practice, as we demonstrate in Section~\ref{sec:numerical}.  

To illustrate this property, consider an alternative scenario where the population counterparts of the candidate directions are not aligned, but instead exhibit a common angular separation $\phi_0$ for all pairs.
%For candidate directions $\tilde\vv_{(k)}^\ast = \hat\Vv_{(k)}\wv_{(k)}^\circ$ ($k\in [K])$, this means that $\phi_0 = \angle (\Vv_{(k)}\wv_{(k)}, \Vv_{(l)}\wv_{(l)})$ for all pairs. 
In this case, it follows from Theorem~\ref{thm:oracle}(a) that the (oracle) standardized statistic converges to
$-{\Delta}^{\text{st}}(\phi_0) = -2\sqrt{{n}/{{\Vc}^\circ}} \sum_{k<l} m_{(k)}m_{(l)}(1 - \cos \phi_0)$. That is, under an alternative with angular separation of at least $\phi_0$, the standardized statistic is expected to be shifted away from the origin by an amount $\Delta^{\text{st}}(\phi_0)$.  
For even moderate sample sizes and signal strengths, this shift is substantial.  For instance, even a small angle such as $\phi_0 = 10^\circ$ typically results in a $\Delta^{\text{st}}(\phi_0)$ far exceeding the threshold $\abs{-6}$, which in turn leads to accurate determination of the shared joint rank.
  
\begin{figure}[t]
    \centering    
    \includegraphics[width=0.9\linewidth]{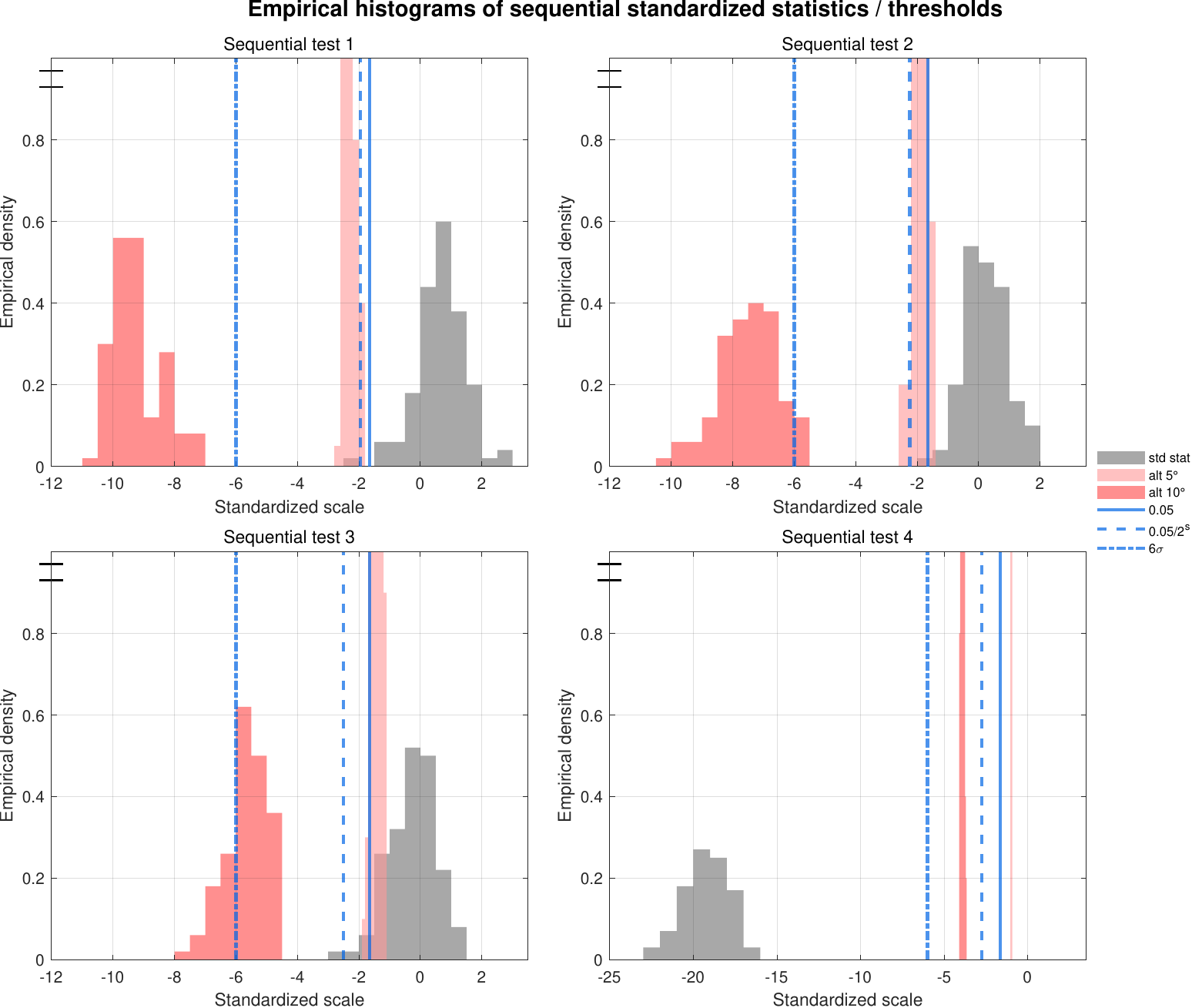}
    \caption{Various thresholds compared with the empirical distributions of $\tilde{Z}$ (gray histogram); the first three panels show the null distribution, and the last panel shows an alternative distribution.  For all instances, utilizing the $6\sigma$ threshold results in $\hat{r}_{\textsc{j}} = 3$, recovering the true rank $r_{\textsc{j}} = 3$. 
    }
    \label{fig:var.thres}
\end{figure}

Figure~\ref{fig:var.thres} evaluates various thresholding strategies using simulated data sets with true joint rank $r_{\textsc{j}} = 3$ (corresponding to the base model discussed in Section~\ref{sec:sim.two_block}). The empirical sampling distributions of the standardized statistics $\tilde{Z}$ \eqref{eq:seqtest_procedure}, corresponding to each of the first four candidate directions $\tilde\vv_{\textsc{j},s}$ ($s= 1,\ldots, 4)$ in the sequential alignment test procedure, are  replicated over 100 repetitions. 

As illustrated in Figure~\ref{fig:var.thres}, both the classical threshold ($-z_{0.05}$) and the FWER-adjusted threshold ($-z_{\alpha_s}$ with $\alpha_s = 0.05/2^s$) lead to few instances of the false rejection of valid joint components. This underestimation persists even in large-sample regimes. In contrast, the $6\sigma$ rule correctly identifies all three joint components for all realizations. The empirical statistics and the threshold in the last panel correctly suggests that the fourth candidate directions are not shared, and our procedure leads to $\tilde{r}_{\textsc{j}} = 3$. 

Figure~\ref{fig:var.thres} also shows the empirical counterpart of the theoretical shift $-{\Delta}^{\text{st}}(\phi_0)$ for $\phi_0 = 5^\circ$ and $10^\circ$, defined as  
\begin{align}\label{eq:alt.cut}
    -\tilde{\Delta}^{\text{st}}(\phi_0) = -2\sqrt{\frac{n}{\tilde{\Vc}^\circ_{\yv}(\lambdav)}} \sum_{k<l} {\hat{m}_{(k)} \hat{m}_{(l)}}(1 - \cos \phi_0),
\end{align}
where $\tilde{\Vc}^\circ_{\yv}(\lambdav)$ and $\hat{m}_{(k)}$ are given in (\ref{eq:working_mean_var}) and
\eqref{eq:k_working_mean_var}, respectively. Since the moment estimators are data-dependent, the empirical shifts vary across realizations.
If a relaxed criterion is preferred to allow for a user-specified angular tolerance $\phi_0$, one could adopt the estimated shift $-\tilde{\Delta}^{\text{st}}(\phi_0)$ as the threshold. 
We do not pursue this approach here, and employ the $6\sigma$ threshold throughout our experiments.

\subsection{Reconstruction of the joint and individual components}\label{sec:reconstruction}

Once the joint rank and score subspaces are identified, our final goal is to reconstruct the joint and individual components for each data block $\Xv_k$.
Our approach is a projection-based decomposition, as similarly done in \cite{feng2018angle, prothero2024data}. 

First, the estimated basis  $\tilde{\Vv}_\Jv$ of the joint row subspace is used to reconstruct the joint component of $\Xv_k$ as  $    \tilde{\Jv}_k := \Xv_k \Pv_{\tilde{\Vv}_{\Jv}}$.
The individual component, representing the source-specific variation, is obtained by extracting the signal that remains after removing the joint-aligned contribution from the total signal subspace. Specifically, we subtract the projection onto the joint-aligned subspace from the $r_k$-dimensional source-specific signal projection:
\begin{align*}
    \tilde{\Iv}_k := \Xv_k \left( \Pv_{\hat{\Vv}_{(k)}} - \Pv_{\hat{\Vv}_{(k)}|\tilde{\Vv}_\Jv} \right), \quad \Pv_{\hat{\Vv}_{(k)}|\tilde{\Vv}_\Jv}  = \Pv_{\hat{\Vv}_{(k)}}\tilde{\Vv}_\Jv \left( \tilde{\Vv}_\Jv^\top\Pv_{\hat{\Vv}_{(k)}}\tilde{\Vv}_\Jv \right)^{-1}\tilde{\Vv}_\Jv^\top \Pv_{\hat{\Vv}_{(k)}}.
\end{align*}
Here, the joint-aligned subspace is spanned by $\Pv_{\hat{\Vv}_{(k)}}\tilde{\Vv}_\Jv$, i.e., the projections of $\tilde{\Vv}_\Jv$ onto the source-specific signal subspace. 

This construction ensures that the resulting row subspaces of $\tilde{\Jv}_k$ and $\tilde{\Iv}_k$ have dimensions $\tilde{r}_{\textsc{j}}$ and $\hat{r}_k - \tilde{r}_{\textsc{j}}$, respectively. Furthermore, the reconstructed components satisfy the structural requirements for integrative multi-source factor models as specified in \eqref{eq:joint_subspace}.

\subsection{Identifying partially shared components}\label{sec:partially.shared}

Up to this point, we focused on estimating the rank of \emph{global} joint components shared across all $K$ blocks. 
However, shared variation is often \textit{partial}, meaning that the variation is common only to a specific subset of the data sources \citep[][see also (\ref{eq:partially_joint})]{gao2021covariate}. To identify such structures, we extend the MSSAT framework into a hierarchical and sequential decomposition procedure.
 
The process begins by identifying and extracting global joint components using the full set of $K$ blocks, using Algorithm \ref{alg:sequential}. Once detected, these global contributions are reconstructed and removed from each block using the projection-based subtraction described in Section~\ref{sec:reconstruction}.
We then decrease the subset size and repeat: for $m=K-1,K-2,\ldots,2$, we apply the sequential alignment tests and reconstruction  to each $m$-block subset (following a fixed ordering) to detect components shared within that subset but not explained by higher-order sharing, and remove them from the participating blocks before proceeding.
The procedure stops at $m=2$, yielding pairwise shared components, and any remaining signal is treated as block-specific individual structure.

The above sequential deflation procedure is order-dependent, as the components removed in earlier stages influence the detection of those in later stages. 
Developing a fully order-invariant extension with formal guarantees on uniqueness and stability remains an important direction for future research.

%===============================================================================

\section{Simulation studies}\label{sec:numerical}
In this section, we evaluate the joint rank estimation accuracy and the quality of the joint-individual decomposition obtained by the proposed method, and compare its performance with several competing approaches. Throughout, our method is referred to as MSSAT, an acronym for \textit{Multi-Source Sequential Alignment Tests}.  

\subsection{Competing methods}

We evaluate the performance of MSSAT against four established approaches for joint--individual decomposition: AJIVE \citep{feng2018angle}, BIDIFAC+ \citep{lock2022bidimensional}, DIVAS \citep{prothero2024data}, and PPD \citep{sergazinov2025spectral}. 

Conceptually, AJIVE, DIVAS, and PPD share a fundamental similarity with MSSAT, as they all distinguish joint components from individual ones based on the principal angles between the estimated score subspaces. Specifically, AJIVE and DIVAS utilize matrix perturbation bounds to establish a lower threshold for these angles. However, because these bounds are inherently conservative and do not account for the exact or asymptotic distributions of empirical angles, they frequently lead to an overestimation of the joint rank. The recently proposed PPD \citep{sergazinov2025spectral} introduces a tighter lower bound. Nevertheless, like AJIVE and DIVAS, PPD relies heavily on computationally expensive resampling strategies to calibrate the threshold.

In contrast, BIDIFAC+ \citep{lock2022bidimensional} formulates the problem through nuclear-norm-regularized matrix factorization, where a structured penalty induces a low-rank decomposition into joint and block-specific components. Although BIDIFAC+ was originally developed for bidimensionally linked matrices sharing both row and column structures, we restrict its application to the special case where blocks share only the row subspace. 
  
Except for BIDIFAC+, all compared methods depend on the accuracy of the initial signal rank estimation. For DIVAS and PPD, we strictly adhere to their respective original protocols for initial rank determination, which are based on the hard thresholding methods of \citet{gavish2014optimal,gavish2017optimal}. For AJIVE, we use our procedure, described in Section~\ref{sec:estimation}, to set  the initial signal ranks. 

\subsection{Two-block cases}\label{sec:sim.two_block}

We first consider the setup with $K = 2$ data blocks, where two distinct data sources are observed for a common set of $n$ individuals. The data are generated according to the integrative multi-source factor model \eqref{model:jive}. We fix the total signal ranks for the two blocks at $r_1 = \operatorname{rank}(\Av_1) = 6$ and $r_2 = \operatorname{rank}(\Av_2) = 9$, respectively, and set the true joint rank to $r_{\textsc{j}} = \operatorname{dim}(\operatorname{row}(\Jv_k)) = 3$. Consequently, the true ranks of the individual components $\Iv_1$ and $\Iv_2$ are $3$ and $6$, respectively.

To evaluate the methods under a challenging geometric configuration, we allow the individual row subspaces to be non-orthogonal. Specifically, all three principal angles between the two individual row subspaces are set to a common value, $\theta_{\mathrm{ind}} = \angle (\operatorname{row}(\Iv_1), \operatorname{row}(\Iv_2)) = \pi/8 = 22.5^\circ$, as the baseline model. We then vary this angle across $\theta_{\mathrm{ind}} \in \{0, \pi/8, \pi/4, 3\pi/8\}$ to assess the impact of subspace alignment. The individual SNRs are set to satisfy $\lambda_{(1),l} \in [3,11]\lambda$ for $l = 1, \dots, 6$ and $\lambda_{(2),l} \in [2,10]\lambda$ for $l =1, \dots, 9$. We choose $\lambda = 1$ as the baseline scale parameter and vary it across $\lambda \in \{0.2, 0.5, 1, 2, 4\}$ to modulate the signal strength. Detailed parameter settings can be found in the Supplementary Material. 

The dimensions and the sample size are fixed at $(n,p_1,p_2) = (500,400,300)$ throughout these experiments. Effects of varying aspect ratios $y_k=p_k/n$  and sample sizes $n$ are given in Table~\xdocref{S1} of the Supplementary Material, in which it is shown that the proposed methods perform well across settings. For each parameter setting, we generate the two-block observation matrices $\Xv_1$ and $\Xv_2$ over 100 independent replication cycles. For each replication, we record the estimated joint rank $\hat{r}_{\textsc{j}}$ and evaluate the quality of the reconstructed joint and individual components, $\tilde{\Jv}_k$ and $\tilde{\Iv}_k$, across all competing methods.

\begin{table}[t]
\centering
\small
\caption{
Simulation results for the two-block baseline model, where the true rank is $r_{\textsc{j}} = 3$. } 
\label{tab:sim_result}
{\fontsize{9pt}{11pt}\selectfont
\begin{tabular}{lccccc}
\toprule
 & \textbf{MSSAT} & \textbf{AJIVE} & \textbf{BIDIFAC+} & \textbf{DIVAS} & \textbf{PPD} \\
\midrule
\multicolumn{6}{c}{\textit{Estimated joint rank (frequency out of 100)}} \\
\midrule
0                & 0   & 0  & 0  & 0   & 0   \\
1                & 0   & 0  & 0  & 0   & 0   \\
2                & 0   & 0  & 0  & 0   & 0   \\
$\boldsymbol{3}$ & \textbf{100} & \textbf{0}  & \textbf{0}  & \textbf{0}   & \textbf{0}  \\
4                & 0   & 0  & 0  & 0   & 87  \\
5                & 0   & 92 & 0  & 54  & 13  \\
6                & 0   & 8  & 88 & 46  & 0   \\
7                & 0   & 0  & 12 & 0   & 0   \\
% 8                & 0   & 0  & 0  & 0  & 0   \\
\midrule
\multicolumn{6}{c}{\textit{Reconstruction error: mean (sd)}} \\
\midrule
$\Jv_1$     & 0.0289 (0.0021) & 0.8127 (0.0058) & 0.2022 (0.0086) & 0.7968 (0.0424) & 0.6468 (0.0452) \\
$\Jv_2$     & 0.0307 (0.0014) & 0.3172 (0.0044) & 0.1774 (0.0038) & 0.2908 (0.0169) & 0.1928 (0.0310) \\
$\Iv_1$& 0.0450 (0.0027) & 0.9782 (0.0115) & 0.3969 (0.0082) & 0.9729 (0.0251) & 0.8271 (0.0495) \\
$\Iv_2$& 0.0384 (0.0009) & 0.1239 (0.0006) & 0.1509 (0.0010) & 0.1238 (0.0006) & 0.0934 (0.0090) \\
\midrule
\multicolumn{6}{c}{\textit{Runtime (seconds)}} \\
\midrule
time & 2.15 & 4.93 & 196.61 & 1441.96 & 191.25 \\
\bottomrule
\end{tabular}
}
\end{table} 

Table~\ref{tab:sim_result} summarizes the empirical performance of the five methods under the baseline setting ($\lambda = 1, \theta_{\mathrm{ind}} = \pi/8$). 
The proposed MSSAT correctly recovers the true joint rank $r_{\textsc{j}}=3$ in all $100$ instances, while competing methods tend to overestimate the joint rank. This is expected, because the competing methods use either conservative bounds, or is developed under the strict assumption of mutually orthogonal individual components. 

This advantage in rank detection directly translates into the superior quality of the joint-individual decomposition achieved by MSSAT. 
To quantify the component-wise recovery, we evaluate the reconstruction accuracy via the relative Frobenius error:
\begin{equation}\label{eq:reconstruct.error}
    \operatorname{Err}(\Sv,\tilde{\Sv})
= \frac{\|\Sv - \tilde{\Sv}\|_F^2}{\|\Sv\|_F^2},
\end{equation}
where $\Sv$ denotes the true component ($\Jv_k$ or $\Iv_k$) and $\tilde{\Sv}$ denotes its corresponding estimate ($\tilde{\Jv}_k$ or $\tilde{\Iv}_k$). As shown in Table~\ref{tab:sim_result}, the reconstruction performance of MSSAT outperforms all other competing methods by a substantial margin. 

Finally, Table~\ref{tab:sim_result} reports  the computation times required for the joint rank estimation and matrix reconstruction. On average, MSSAT requires only $2.15$ seconds to complete the process. In stark contrast, BIDIFAC+ and PPD demand over $3$ minutes, while DIVAS exceeds $24$ minutes, since these methods invoke either intensive iterative optimization procedures or time-consuming resampling methods.

\begin{figure}[t]
    \centering
    \includegraphics[width=1\linewidth]{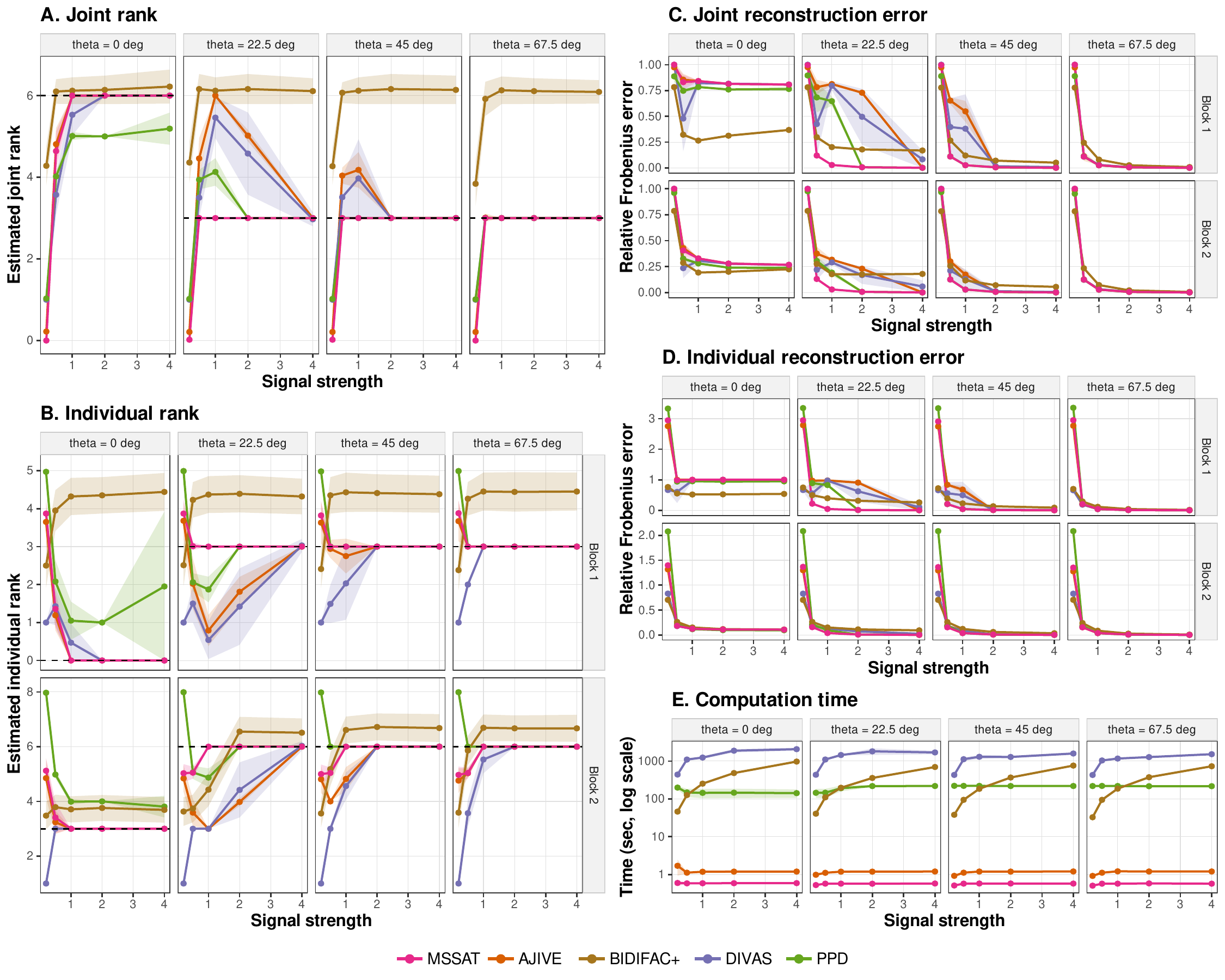}
    \caption{Two-block simulation results as a function of signal strength $\lambda \in \{0.2, 0.5, 1, 2, 4\}$ and individual-subspace angle $\theta_{\mathrm{ind}}\in\{0,\pi/8,\pi/4,3\pi/8\}$.
    Panels (A)--(E) summarize rank estimates, reconstruction errors, and average computation times for MSSAT and competing methods.}
    \label{fig:sim_d_ang}
\end{figure}

Simulation results for varying SNR scale parameter $\lambda$ and varying principal angles $\theta_{\mathrm{ind}}$ between individual row subspaces are summarized in Figure \ref{fig:sim_d_ang}. Joint component segmentation becomes more challenging as the principal angles diminish and the signal strength weakens. Panels (A) and (B) report the estimated joint and individual ranks, panels (C) and (D) present the component-wise reconstruction errors (\ref{eq:reconstruct.error}), and panel (E) displays the average computation time on a log scale. 

Note that when $\theta_{\mathrm{ind}}  =0$, the individual subspaces collapse into each other, and the true joint rank becomes $r_{\textsc{j}} = 6$ (and reducing the individual ranks to be $0$ and $3$, respectively), instead of the default $r_{\textsc{j}} = 3$.  the proposed MSSAT outperforms all competing methods, with its superiority being most pronounced in regimes with moderate signal strengths and smaller principal angles. Notably, subspace-based methods, including MSSAT, AJIVE, DIVAS, and PPD, which leverage the alignment of block-wise score subspaces, converge to the correct target as the signal strength grows. In contrast, BIDIFAC+  overestimates the joint rank across nearly all settings, because its nuclear-norm penalization fails to decouple individual components when their scores are correlated.

\subsection{Multi-block cases}\label{sec:sim.multi}

We next investigate the performance of MSSAT in a multi-block setting involving $K = 3$ data blocks. When data strictly follow the standard integrative factor model (\ref{model:jive}) with only \textit{globally shared} joint components, the relative performance advantages of MSSAT are qualitatively identical to those observed in the $K=2$ case. 
 Therefore we instead consider a generalized framework that simultaneously incorporates both globally and \textit{partially shared} structures among the three data sources. Specifically, we generate the data according to the following partially-shared integrative factor model:
\begin{equation}
    \begin{aligned}
            \Xv_1 &= \Jv_1 + \Jv_{1}^{\{1,2\}} + \Jv_{1}^{\{1,3\}} \quad\qquad\,\,\, + \Iv_1 + \Ev_1, \\ 
    \Xv_2 & = \Jv_2 + \Jv_{2}^{\{1,2\}} \quad\qquad\,\,\, + \Jv_{2}^{\{2,3\}} + \Iv_2 + \Ev_2, \\
    \Xv_3 & = \Jv_3 \quad\qquad\,\,\, + \Jv_{3}^{\{1,3\}} + \Jv_{3}^{\{2,3\}} + \Iv_3 + \Ev_3, 
    \end{aligned}
    \label{eq:partially_joint}
\end{equation} 
where the globally shared joint component $\Jv_k$ and the individual component $\Iv_k$ satisfy the requirement (\ref{eq:joint_subspace}). 
For each pair, the partially shared components $\Jv_{k}^{\{k,\ell\}}$ share the row score subspaces, i.e., $\text{row}(\Jv_{k}^{\{k,\ell\}}) = \text{row}(\Jv_{\ell}^{\{k,\ell\}})$, which are perpendicular to other pairs, as well as to the globally joint and individual row subspaces. 

We note that BIDIFAC+ and DIVAS are natively designed to identify and segment both globally and partially shared joint components. In contrast, AJIVE and PPD are originally formulated to capture only globally shared structures. To ensure a comprehensive evaluation, these latter two methods, along with the proposed MSSAT, are adapted to estimate partially joint components by sequentially utilizing the sub-block extraction strategy described in Section~\ref{sec:partially.shared}.

For simulation experiments, we fix the global joint rank at $r_{\textsc{j}} =\text{dim}(\text{row}(\Jv_k))= 3$, and set the ranks of the partially shared components to $\{r_{12}, r_{13}, r_{23}\} = \{1,1,2\}$, where $r_{k\ell} =\text{dim}(\text{row}(\Jv_{k}^{\{k,\ell\}}))$. The block-wise individual components $\Iv_1, \Iv_2$ and $\Iv_3$ have ranks equal to 1, 3, and 6, respectively. Similar to the settings in the $K=2$ case, the individual scores of Blocks 2 and 3 are set to be moderately correlated with each other, with all three principal angles set to be  $\pi/8$. Other pairs of individual scores are set to be uncorrelated. Details of the parameter setting can be found in the Supplementary Material. 

Throughout these experiments, the sample size and dimensions of the data blocks are fixed at $(n,p_1,p_2,p_3) = (500, 400, 300, 200)$. Table~\ref{tab:sim_result_multi} reports the resulting estimated ranks and component-wise reconstruction errors averaged across 100  replications.

\begin{table}[t]
\centering
\small
\caption{Multi-block simulation results comparing five methods in a three-block partially joint model.
}
\label{tab:sim_result_multi}
{\fontsize{8pt}{11pt}\selectfont
\begin{tabular}{lcccccc}
\toprule
 & \textbf{True rank} & \textbf{MSSAT} & \textbf{AJIVE} & \textbf{BIDIFAC+} & \textbf{DIVAS} & \textbf{PPD} \\
\midrule
\multicolumn{7}{c}{\textit{Mean estimated joint rank}} \\
\midrule
Global      & 3 & 3  & 3    & 3.98  & 3     & 4 \\
Joint12     & 1 & 1  & 1    & 1.10  & 1     & 0.99 \\
Joint13     & 1 & 1  & 1    & 4.57  & 1     & 1 \\
Joint23     & 2 & 2  & 5    & 7.45  & 4.59  & 3.13 \\
Individual1 & 1 & 1  & 1    & 1.64  & 1     & 0.01 \\
Individual2 & 3 & 3  & 0    & 1.00  & 0     & 0.88 \\
Individual3 & 6 & 6  & 3    & 11.90 & 3.41  & 3.87 \\
\midrule
\multicolumn{7}{c}{\textit{Reconstruction error: mean (sd)}} \\
\midrule
Global      &  & 0.0166 (0.0004) & 0.0166 (0.0004) & 0.2198 (0.0095) & 0.0433 (0.0013) & 0.2089 (0.1276) \\
Joint12     &  & 0.0147 (0.0005) & 0.0149 (0.0007) & 0.0376 (0.0012) & 0.0205 (0.0025) & 0.4217 (0.3849) \\
Joint13     &  & 0.0375 (0.0020) & 0.0382 (0.0021) & 0.5301 (0.0476) & 0.1317 (0.0263) & 0.1373 (0.1337) \\
Joint23     &  & 0.0110 (0.0005) & 0.3313 (0.0027) & 0.1085 (0.0046) & 0.3476 (0.0328) & 0.1863 (0.0498) \\
Individual1 &  & 0.2724 (0.1255) & 0.2265 (0.0095) & 0.3797 (0.0135) & 0.2807 (0.0366) & 0.9925 (0.0747) \\
Individual2 &  & 0.1967 (0.0219) & 1.0000 (0.0000) & 0.9991 (0.0008) & 1.0000 (0.0000) & 0.8417 (0.1341) \\
Individual3 &  & 0.0091 (0.0008) & 0.0660 (0.0005) & 0.1758 (0.0038) & 0.0684 (0.0005) & 0.0576 (0.0156) \\
\midrule
\multicolumn{7}{c}{\textit{Runtime (seconds)}} \\
\midrule
time &  & 5.62 & 10.42 & 1994.63 & 6006.20 & 774.04 \\
\bottomrule
\end{tabular}
}
\end{table}

MSSAT accurately recovers the global, pairwise partially shared, and block-specific individual ranks across all instances. In contrast, competing methods tend to exhibit overestimation of joint ranks. For example, all competing methods overestimate the Block 2–3 pairwise rank (due to the correlated individual scores), and PPD tend to overestimate the global joint rank as well. BIDIFAC+ suffers from most pervasive overestimation.  
The component-wise reconstruction errors directly mirror these rank estimation patterns; %As seen in Table~\ref{tab:sim_result_multi}, 
MSSAT achieves lowest errors across all structural components

\section{Real data analysis}\label{sec:realdata}

\subsection{TCGA Breast Cancer Dataset}

We demonstrate the practical utility of MSSAT on a multi-omics dataset from The Cancer Genome Atlas (TCGA), which has served as a canonical benchmark for joint--individual decomposition \citep{feng2018angle}. 
Specifically, we analyze the $n = 616$ breast tumor samples studied by \citet{ciriello2015comprehensive}. Each sample is measured across four heterogeneous genomic platforms: 
1,615 gene expression (GE) features, 24,174 copy number variation (CN) features, 187 reverse-phase protein array (RPPA) features, and 18,256 mutation (MUT) features.
After preprocessing the breast cancer data set as done in \citet{feng2018angle},
we first apply the proposed MSSAT, as well as AJIVE, BIDIFAC+, DIVAS, and PPD, to determine whether the four data blocks share any statistically significant joint score directions (i.e., globally shared components) across different omics platforms.

\paragraph{Globally shared components}
Table~\ref{tab:TCGA} reports the estimated global joint rank and block-wise individual ranks, together with their computation times, for MSSAT and competing methods on the TCGA breast cancer multi-omics data. 
Both MSSAT and PPD find no statistical evidence of a common score subspace shared across all four heterogeneous omics blocks, yielding $\hat{r}_{\textsc{j}} = 0$. In contrast, AJIVE, BIDIFAC+, and DIVAS estimate nonzero global joint ranks of $26$, $1$, and $2$, respectively.  

\begin{table}[t]
\centering
\small
\caption{TCGA breast cancer multi-omics analysis comparing five methods. 
The table reports the estimated global joint rank shared across all four blocks (GE, CN, RPPA, MUT), block-specific (individual) ranks for each omics type, and total runtime (in seconds) for each method.}
\label{tab:TCGA}
\begin{tabular}{lccccc}
\toprule
 & \textbf{MSSAT} & \textbf{AJIVE} & \textbf{BIDIFAC+} & \textbf{DIVAS} & \textbf{PPD} \\
\midrule
\multicolumn{6}{c}{\textit{Estimated rank}} \\
\midrule
Joint & 0  & 26  & 1 & 2  & 0  \\
$\textnormal{GE}_{ind}$   & 180  & 154  & 124 & 50   & 125  \\
$\textnormal{CN}_{ind}$   & 215  & 189  & 21  & 111  & 219  \\
$\textnormal{RPPA}_{ind}$ & 41   & 15   & 23  & 11   & 47  \\
$\textnormal{MUT}_{ind}$  & 183  & 157  & 1   & 71   & 116  \\

\midrule
\multicolumn{6}{c}{\textit{Runtime (seconds)}} \\
\midrule
time & 177 & 2,074 & 206,589 & 20,376 & 111,937 \\
\bottomrule
\end{tabular}
\end{table}

This discrepancy aligns with the theoretical characteristics of each method and echoes the overestimation patterns observed in our simulation studies (Section~\ref{sec:numerical}). Because AJIVE and DIVAS rely on conservative perturbation-based bounds, they are highly prone to capturing weak or partially aligned variations as global joint components; in Table~\ref{tab:TCGA}, this appears as AJIVE's markedly inflated joint rank of $26$. Overall, under a strict notion of global sharing across all four omics platforms, MSSAT yields a cautious estimate ($\hat{r}_{\textsc{j}}=0$) and successfully avoids the joint-rank inflation induced by conservative angle bounds.

Table~\ref{tab:TCGA} also shows that MSSAT runs substantially faster the the competing methods. For this ultra-high dimensional data set, MSSAT completes the decomposition in just 177 seconds (less than 3 minutes).

\paragraph{Partially shared components} 
Having examined globally shared structures, we now turn to identify partially shared components across specific subsets of data blocks. Table~\ref{tab:TCGA_partial} reports the resulting rank estimates for the global and partial joint components,  as well as the block-wise individual components. 
Here, we omit the standard versions of AJIVE and PPD, as their original formulations do not inherently account for hierarchical or partially shared structures.

\begin{table}[t]
\centering
\caption{Estimated ranks of (partially) shared components for the TCGA breast cancer multi-omics data.
For each subset of $\{$GE, CN, RPPA, MUT$\}$, the table reports the dimensions of score subspaces identified as common to the blocks in that subset. 
}
\label{tab:TCGA_partial}
\begin{tabular}{llccc}
\toprule
Size & Combination 
& MSSAT &  BIDIFAC+ & DIVAS \\
\midrule
4
 & $\{\mathrm{GE},\mathrm{CN},\mathrm{RPPA},\mathrm{MUT}\}$ & 0  & 1 & 2 \\
\midrule
3
 & $\{\mathrm{GE},\mathrm{CN},\mathrm{RPPA}\}$  & 0  & 4 & 6 \\
 & $\{\mathrm{GE},\mathrm{CN},\mathrm{MUT}\}$   & 0  & 0 & 9 \\
 & $\{\mathrm{GE},\mathrm{RPPA},\mathrm{MUT}\}$ & 0  & 1 & 0 \\
 & $\{\mathrm{CN},\mathrm{RPPA},\mathrm{MUT}\}$ & 0  & 0 & 0 \\
\midrule
2
 & $\{\mathrm{GE},\mathrm{CN}\}$    & 0  & 13 & 47 \\
 & $\{\mathrm{GE},\mathrm{RPPA}\}$  & 0   & 16 & 9  \\
 & $\{\mathrm{GE},\mathrm{MUT}\}$   & 1   & 0  & 13 \\
 & $\{\mathrm{CN},\mathrm{RPPA}\}$  & 0   & 1  & 4  \\
 & $\{\mathrm{CN},\mathrm{MUT}\}$   & 7  & 0  & 34 \\
 & $\{\mathrm{RPPA},\mathrm{MUT}\}$ & 0   & 0  & 1  \\
\midrule
1
 & $\{\mathrm{GE}\}$   & 179  & 90  & 50  \\
 & $\{\mathrm{CN}\}$   & 208  & 3   & 111 \\
 & $\{\mathrm{RPPA}\}$ & 41   & 1   & 11  \\
 & $\{\mathrm{MUT}\}$  & 174  & 0   & 71  \\
\midrule
\multicolumn{5}{c}{\textit{Runtime (seconds)}} \\
\midrule
time &  & 972   & 206,589 & 20,376 \\
\bottomrule
\end{tabular}
\end{table}

Consistent with the global analysis, MSSAT finds no joint components shared across any three-block subset, supporting the conclusion that globally coherent structure spanning all omics platforms is weak in this dataset. In contrast, BIDIFAC+ and DIVAS report nontrivial joint ranks for several multi-block combinations. 
In the following, we carefully analyze the joint components identified by DIVAS to demonstrate that their cross-block overlaps are excessively weak, while the partially joint components identified by MSSAT are geometrically plausible. 

Specifically, we evaluate the extent to which the reconstructed joint component in each block lies outside that block’s estimated initial signal row subspace. 
For each block $k$, define the relative projection error
\begin{align*} \operatorname{Err}\!\left(\hat{\Pv}_{(k)}, \tilde{\Jv}_k\right)
    := \frac{\big\|\big(\Iv_n-\hat{\Pv}_{(k)}\big)\tilde{\Jv}_k\big\|_F^2}{\big\|\tilde{\Jv}_k\big\|_F^2},
\end{align*}
where $\hat{\Pv}_{(k)}=\hat{\Vv}_{(k)}\hat{\Vv}_{(k)}^\top$ is the orthogonal projector onto $\operatorname{span}(\hat{\Vv}_{(k)})$ and $\tilde{\Jv}_k$ is the reconstructed joint component for block $k$.
Smaller values indicate that $\tilde{\Jv}_k$ is largely contained in $\operatorname{span}(\hat{\Vv}_{(k)})$, whereas larger values indicate weaker cross-block coherence.

Applying this diagnostic to the DIVAS global joint estimate, we project it onto each block’s initial low-rank signal row subspace
(with ranks $180$ (GE), $215$ (CN), $41$ (RPPA), and $183$ (MUT)) and obtain
$\operatorname{Err}=0.22$ (GE), $0.60$ (CN), $0.63$ (RPPA), and $0.54$ (MUT).
Thus, a substantial fraction of the DIVAS' global joint variation stems from noise outside the block-specific signal subspaces.
An identical conclusion is reached when comparing the row subspaces directly: the largest principal angles between the DIVAS global joint subspace ($\operatorname{span}(\tilde{\Vv}_\Jv)$) and the block-wise signal row subspaces ($\operatorname{span}(\hat{\Vv}_{(k)})$) are
$58.11^{\circ}$ (GE), $56.63^{\circ}$ (CN), $62.36^{\circ}$ (RPPA), and $57.82^{\circ}$ (MUT).

In contrast, for the shared subspaces retained by MSSAT in this dataset, both diagnostics are much smaller.
For the GE--MUT shared space we obtain $\operatorname{Err}=0.01$ and a largest principal angle of $6.05^{\circ}$, and for the CN--MUT shared space we obtain $\operatorname{Err}=0.007$ and $5.31^{\circ}$, indicating substantially tighter cross-block alignment under MSSAT.

\subsection{Korean mortality data}

We demonstrate an application of MSSAT to Korean mortality data from 1969 to 2023, successfully decoupling macro-level public health trends from gender-specific socio-behavioral dynamics. The age–specific mortality data from the Korean population are obtained for each of male and female cohorts. 
For each gender, we observe a matrix of $\log_{10}$ transformed mortality rates, with rows corresponding to ages 0 through 100 and columns representing 55 calendar years from 1969 to 2023. 
Because mortality levels vary substantially across ages, all data matrices are row-centered so that broad age wide level differences, which affect all years uniformly, are removed prior to analysis.

After preprocessing, the two blocks, male and female mortality, form a pair of aligned datasets with identical sample indices (years) but potentially different age-specific structures. We directly apply MSSAT to these matrices to identify any low-dimensional structure that may be shared across genders, as well as to separate block-specific patterns unique to male or female mortality.

\begin{figure}[t]
    \centering
    \includegraphics[width=1\linewidth]{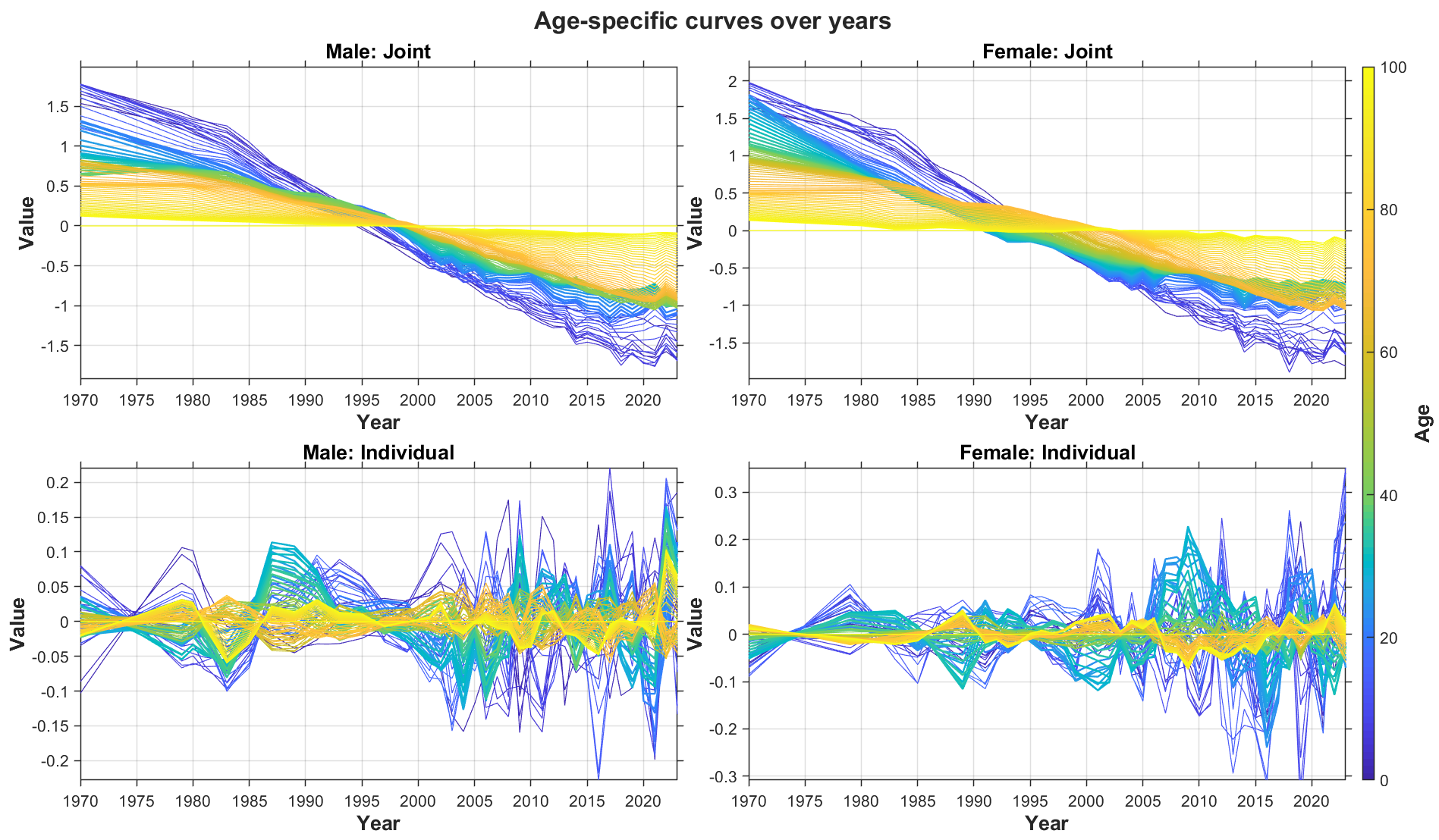}
   \caption{Decomposition of Korean mortality data into joint and individual components using MSSAT. Top: Estimated joint structures shared by male (left) and female (right) mortality. Bottom: Corresponding gender-specific individual structures. Each curve represents an age-specific mortality trajectory (ages 0–100) over time, colored by age.}
    \label{fig:kor_mortality}
\end{figure}

Figure~\ref{fig:kor_mortality} displays the decomposition of the Korean mortality data into joint and individual structures obtained by MSSAT (see also Supplementary Figure~\xdocref{S1} for matrix visualizations). The estimated global joint rank is $\tilde{r}_{\textsc{j}}=5$, while each gender-specific individual component has a rank of $6$. 
The joint components capture common temporal downward trends across genders, whereas the individual components highlight distinct, gender-specific deviations from the shared baseline.

The joint components capture a pronounced long-term downward trend in mortality for both genders, reflecting macroeconomic and public health improvements in Korea. This decline is steepest at younger ages, driven by well-documented reductions in infant and child mortality since the 1970s. Additionally, a slight uptick emerges around 2020, plausibly capturing the temporary impact of the COVID-19 pandemic on all-cause mortality.

In contrast, the individual components exhibit distinct gender-specific patterns of variation. While these components account for a smaller proportion of the total variation, some of their most pronounced patterns closely align with known demographic facts. For example, the excess mortality among young men in the 1990s (indicated by the teal-colored peak) is historically attributed to an abrupt rise in external causes of death, most notably soaring traffic accident fatalities during a period of rapid motorization \citep{lim2014trends}. On the other hand, the delayed mortality peak among young women in the late 2000s may reflect an era of shifting social dynamics, linked to increased socioeconomic stress and rising suicide rates among young women during that period \citep{park2018age}. While causal demographic analysis is beyond our scope, the proposed method provides a powerful tool to successfully separate shared joint patterns from source-specific distinct variations.

\section{Conclusions}
In this paper, we proposed the Multi-Source Sequential Alignment Test (MSSAT), a novel method for estimating joint structures in multi-source data based on sequential alignment tests for score subspaces. Because the proposed test relies on the asymptotic null distribution of the alignment statistic, it is highly computationally efficient. Furthermore, by carefully choosing the critical threshold, MSSAT is shown to achieve highly accurate joint rank estimation.

Despite these advantages, certain limitations warrant further investigation. First, our theoretical development relies on an idealized setting where the joint structure corresponds to identical singular directions across data blocks. In practice, however, shared structures may manifest through the overlap of singular subspaces rather than the exact coincidence of singular vectors. While MSSAT can be applied heuristically to such cases by aggregating multiple directions, a rigorous theoretical treatment of this subspace-level alignment remains open. Second, regarding partially shared structures, our numerical experiments demonstrated that MSSAT can be adapted using a greedy hierarchical deflation approach. However, a formally organized methodology and a comprehensive theoretical justification for this specific extension have yet to be fully developed.

\section*{Acknowledgments}

This work was supported by the Samsung Science and Technology Foundation (SSTF-BA2002-03) and the National Research Foundation of Korea (RS-2023-00301976, RS-2024-00333399, and RS-2025-25436116), including the Basic Science Research Program funded by the Ministry of Education.

\bibliographystyle{asa}
\bibliography{mssat}

\end{document}

\title{Supplementary Information for\\
Testing and segmentation of joint and individual components in integrative multi-source factor models}

\author[1]{Kyoowon Kim}
\author[2]{Sungkyu Jung}

\affil[1]{Department of Statistics, Seoul National University, kwkim1224@snu.ac.kr}
\affil[2]{Department of Statistics and Institute for Data Innovation in Science, Seoul National University, sungkyu@snu.ac.kr}

\date{}
\maketitle

\renewcommand{\thesection}{S\arabic{section}}
\renewcommand{\thefigure}{S\arabic{figure}}
\renewcommand{\thetable}{S\arabic{table}}
\renewcommand{\thethm}{S\arabic{thm}}

\setcounter{section}{0}

\tableofcontents

\section{Simulation settings and supplementary numerical results}\label{app:num}

\subsection{Simulation setting for Figures~\xdocref{2} and \xdocref{3}}\label{app:sim.set.2}

\paragraph{Simulation Setting:}
We consider a two-block data-generating model (Figure~\xdocref{2}) and a three-block data-generating model (Figure~\xdocref{3}). 
Both models follow \eqref{eq:model.sim}.

\paragraph{Full Model.}
For $k=1,2,3$,
\begin{align}\label{eq:model.sim}
    \Xv_k = d_{(k),1}\, \uv_{(k),1} \vv_{\textsc{j}}^\top + \sum_{\ell=2}^{r_k}
d_{(k),\ell} \uv_{(k),\ell} \vv_{(k),\ell}^\top + \Ev_k.
\end{align}
Thus each block consists of a shared rank-one joint component and block-specific individual components with heterogeneous signal magnitudes and noise levels.

\paragraph{Dimensions.}
Let the sample size be $n = 1000$. The three data blocks are
\[
X_k \in \mathbb{R}^{p_k \times n}, \qquad k = 1,2,3,
\]
with feature dimensions $(p_1,p_2,p_3) = (800,\,600,\,500)$.
The total ranks of the blocks are $(r_1, r_2, r_3) = (2,3,3)$ with joint rank $r_{\textsc{j}}=1$.

\paragraph{Signal Strengths.}
Let $d_{(k)} = (d_{(k),1},\dots,d_{(k),r_k})$ denote the singular values of block $k$:
\[
\begin{aligned}
(d_{(1),1}, d_{(1),2})              &= \sqrt{n}\cdot (2,1)\cdot 10, \\
(d_{(2),1}, d_{(2),2}, d_{(2),3}) &= \sqrt{n}\cdot (1,3,2)\cdot 8, \\
(d_{(3),1}, d_{(3),2}, d_{(3),3}) &= \sqrt{n}\cdot (3,2,1)\cdot 6.
\end{aligned}
\]

For each $k=1,2,3$, the noise matrices satisfy \((\Ev_k)_{i,j} \overset{iid}{\sim} \mathcal{N}(0,\sigma_k^2),\) with \(\sigma_1=\sigma_2=\sigma_3=1.\)
\newline

For Figure~\xdocref{2}, only the first two data blocks, $\Xv_1$ and $\Xv_2$, are used, whereas all three blocks are included in Figure~\xdocref{3}. 
In both figures, 100 replicates are performed to construct the empirical distributions. 
For the alternative scenario in Figure~\xdocref{3}, the angles are set to $\phi_{\ell_1,\ell_2} = \phi_{\ell_1,\ell_3} = \phi_{\ell_2,\ell_3} = 5^\circ$.

\subsection{Simulation setting for Section~\xdocref{5}}\label{sup:sim.set}

\subsubsection{Two block cases}

%%%%%%%%%%%%%%%%%

\paragraph{Simulation Setting: Two-Block Joint and Individual Structure.}
We consider a two-block data-generating model with a rank-$r_{\textsc{j}}$ component shared between both blocks and block-specific individual components. 

\paragraph{Full Model.}
For $k=1,2$, each data block is generated as
\[
\begin{aligned}
\Xv_1 &= \Uv_{\Jv,1}\Dv_{\Jv,1}\Vv_{\Jv,1}^\top + \Uv_{\Iv,1}\Dv_{\Iv,1}\Vv_{\Iv,1}^\top + \Ev_1,\\
\Xv_2 &= \Uv_{\Jv,2}\Dv_{\Jv,2}\Vv_{\Jv,2}^\top + \Uv_{\Iv,2}\Dv_{\Iv,2}\Vv_{\Iv,2}^\top + \Ev_2,
\end{aligned}
\]
where 
\begin{itemize}
    \item $\operatorname{span}(\Vv_{\Jv,1}) = \operatorname{span}(\Vv_{\Jv,2})$ is the shared row subspace of rank $r_{\textsc{j}}$,
    \item $\operatorname{span}(\Vv_{\Iv,1})$ and $\operatorname{span}(\Vv_{\Iv,2})$ are block-specific row subspaces of ranks $r_1 - r_{\textsc{j}}$ and $r_2 - r_{\textsc{j}}$, respectively.
\end{itemize}

\paragraph{Dimensions.}
The sample size and feature dimensions are
\[
n = 500, \qquad (p_1, p_2) = (400, 300),
\]
with total ranks $(r_1, r_2) = (6, 9)$ and $r_{\textsc{j}}=3$ for the shared component.

\paragraph{Signal Strengths.}
% values are scaled with $\sqrt{n}$ and modulated by a signal strength factor $d$. Specifically, the
The singular diagonal matrices of singular values are
\begin{align}
\operatorname{diag}(\Dv_{\Jv,1} \oplus \Dv_{\Iv,1}) &= 1.5 \sqrt{n} \,(6,5,4,7,3,2), \label{eq:sim.signal.1}  \\
\operatorname{diag}(\Dv_{\Jv,2} \oplus \Dv_{\Iv,2}) &=  \sqrt{n} \,(7,6,5,8,9,10,2,3,4), \label{eq:sim.signal.2}
\end{align}

where $\oplus$ denotes the block-diagonal concatenation of component singular values, and $\operatorname{diag}$ extracts the diagonal entries.

\paragraph{Individual Subspace Angle.}
The two data blocks have 3 and 6 individual components, respectively, and the corresponding individual row subspaces are set such that all three principal angles between them are $\pi/8$ ($22.5^\circ$). 
\vspace{10pt}

Table~\xdocref{1} is made with 100 times replication with simulation setting above. 
Figure~\xdocref{5} reports a two-block sensitivity experiment where we vary both the signal-to-noise ratio (SNR) and the geometry of the individual structures.
Specifically, we consider SNR levels 
$\lambda\in\{0.2,0.5,1,2,4\}$ 
by proportionally scaling the population signal singular values, and we control the alignment between the two individual row subspaces via a principal-angle parameter $\phi_{\mathrm{ind}}\in\{0,\pi/8,\pi/4,3\pi/8\}$.
Here, the baseline SNR level $\lambda=1.00$ coincides with the simulation setting used for Table~\xdocref{1}.
The remaining SNR configurations $\lambda\in\{0.2,0.5,1,2,4\}$ are generated by multiplying the baseline population signal singular values by the factor $\lambda$, while keeping the noise variance fixed.
In this design, the three principal angles between the individual row subspaces are all set equal to $\phi_{\mathrm{ind}}$ (so $\phi_{\mathrm{ind}}=0$ corresponds to coincident individual subspaces).

\paragraph{Effects of vaying aspect ratios and sample sizes}
Table~\ref{tab:sim_result_ms_sat} provides an additional robustness check under varying sample sizes $n\in\{100,200,400\}$ and aspect ratios $y_k=p_k/n\in\{0.25,0.50,1.00,2.00,4.00\}$.
The underlying signal construction (i.e., the joint and individual subspaces and the signal strengths) is held fixed, while $(p_k,n)$ are adjusted to achieve the target values of $y_k$ and $n$.
The full model and signal strength settings are identical to those used for Table~\xdocref{1}, except that the individual subspaces are set to be orthogonal.
Each setting in Table~\ref{tab:sim_result_ms_sat} is evaluated based on 100 replications.
MSSAT successfully recovers the true joint rank across almost all configurations, demonstrating robustness to changes in sample size and aspect ratio. 
A slight degradation is observed only in the extremely low-dimensional setting with both small sample size and small aspect ratio.

\begin{table}[t]
\centering
\footnotesize
\caption{Joint-rank estimation performance of MSSAT under varying sample sizes $n$ and aspect ratios $y_k = p_k/n$.
The table reports the mean estimated joint rank together with the proportion of replications in which the true joint rank ($r_{\textsc{j}}=3$) is correctly recovered (in parentheses).}
\label{tab:sim_result_ms_sat}
\begin{tabular}{l>{\centering\arraybackslash}p{2.2cm}
                  >{\centering\arraybackslash}p{2.2cm}
                  >{\centering\arraybackslash}p{2.2cm}}
\toprule
 & \multicolumn{3}{c}{\textbf{MSSAT}} \\
\cmidrule(lr){2-4}
Aspect ratio $y_k$
 & $n=100$ & $n=200$ & $n=400$ \\
\midrule
 & \multicolumn{3}{c}{\textit{Mean estimated joint rank (Proportion accurately estimated)}} \\
\midrule
0.25 & 2.00(0.00) & 3.00(1.00) & 3.00(1.00) \\
0.50 & 3.00(1.00) & 3.00(1.00) & 3.00(1.00) \\
1.00 & 3.00(1.00) & 3.00(1.00) & 3.00(1.00) \\
2.00 & 3.00(1.00) & 3.00(1.00) & 3.00(1.00) \\
4.00 & 3.00(1.00) & 3.00(1.00) & 3.00(1.00) \\
\bottomrule
\end{tabular}
\end{table}

\subsubsection{Multi-block cases}

\paragraph{Simulation Setting: Three-Block Multi-Level Joint Structure.}
We consider a three-block data-generating model with hierarchical low-rank structures: 
a shared global component across all blocks, pairwise joint components for each block pair, and block-specific individual components.

\paragraph{Full Model.}
For $k=1,2,3$,
\begin{align*}
    \Xv_1 &= \Uv_{\Jv,1}\Dv_{\Jv,1}\Vv_{\Jv,1}^\top + \Uv_{\Jv_{12},1}\Dv_{\Jv_{12},1}\Vv_{\Jv_{12},1}^\top + \Uv_{\Jv_{13},1}\Dv_{\Jv_{13},1}\Vv_{\Jv_{13},1}^\top +
    \Uv_{\Iv,1}\Dv_{\Iv,1}\Vv_{\Iv,1}^\top + \Ev_1,\\
    \Xv_2 &= \Uv_{\Jv,2}\Dv_{\Jv,2}\Vv_{\Jv,2}^\top + \Uv_{\Jv_{12},2}\Dv_{\Jv_{12},2}\Vv_{\Jv_{12},2}^\top + \Uv_{\Jv_{23},2}\Dv_{\Jv_{23},2}\Vv_{\Jv_{23},2}^\top +
    \Uv_{\Iv,2}\Dv_{\Iv,2}\Vv_{\Iv,2}^\top + \Ev_2,\\
    \Xv_3 &= \Uv_{\Jv,3}\Dv_{\Jv,3}\Vv_{\Jv,3}^\top + \Uv_{\Jv_{13},3}\Dv_{\Jv_{13},3}\Vv_{\Jv_{13},3}^\top + \Uv_{\Jv_{23},3}\Dv_{\Jv_{23},3}\Vv_{\Jv_{23},3}^\top +
    \Uv_{\Iv,3}\Dv_{\Iv,3}\Vv_{\Iv,3}^\top + \Ev_3,
\end{align*}
where 
\begin{itemize}
    \item $\operatorname{span}(\Vv_{\Jv,1}) = \operatorname{span}(\Vv_{\Jv,2}) = \operatorname{span}(\Vv_{\Jv,3})$ represents the globally shared row subspace, 
    \item $\operatorname{span}(\Vv_{\Jv_{12},1}) = \operatorname{span}(\Vv_{\Jv_{12},2})$, 
          $\operatorname{span}(\Vv_{\Jv_{13},1}) = \operatorname{span}(\Vv_{\Jv_{13},3})$, 
          $\operatorname{span}(\Vv_{\Jv_{23},2}) = \operatorname{span}(\Vv_{\Jv_{23},3})$ are partially shared row subspaces between block pairs,
    \item $\operatorname{span}(\Vv_{\Iv,k})$ is specific to block $k$.
\end{itemize}

\paragraph{Dimensions.}
The sample size is $n=500$, and the three data blocks have feature dimensions
\[
(p_1, p_2, p_3) = (400,\,300,\,200),
\]
with total ranks $(r_1, r_2, r_3) = (6,9,12)$.

\paragraph{Global and Pairwise Joint Components.}
Let $r_{\textsc{j}} = 3$ denote the rank of the global (all-block) component, and let $r_{12}=1$, $r_{13}=1$, $r_{23}=2$ denote the ranks of the pairwise components shared between blocks $(1,2)$, $(1,3)$, and $(2,3)$, respectively.  
The remaining components for blocks 1, 2, and 3 are block-specific individual components with ranks 1, 3, and 6, respectively.  
Among these, the block-specific row subspaces $\operatorname{span}(\Vv_{\Iv,2})$ and $\operatorname{span}(\Vv_{\Iv,3})$ are set such that all three principal angles between them are $\pi/8$ ($22.5^\circ$).

\paragraph{Signal Strengths.}
The singular values for each block are assigned according to the global, pairwise, and block-specific components. Specifically, the diagonal matrices of singular values for block 1, 2, and 3 are given by

\[
\begin{aligned}
\operatorname{diag}(\Dv_{\Jv,1}\oplus \Dv_{\Jv_{12},1}\oplus \Dv_{\Jv_{13},1}\oplus \Dv_{\Iv,1}) &= 1.5\sqrt{n}\,(6,5,4,7,3,2), \\
\operatorname{diag}(\Dv_{\Jv,2}\oplus \Dv_{\Jv_{12},2}\oplus \Dv_{\Jv_{23},2}\oplus \Dv_{\Iv,2}) &= \sqrt{n}\,(7,6,5,8,9,10,2,3,4), \\
\operatorname{diag}(\Dv_{\Jv,3}\oplus \Dv_{\Jv_{13},3}\oplus \Dv_{\Jv_{23},3}\oplus \Dv_{\Iv,3}) &= 2\sqrt{n}\,(10,9,8,7,5,6,2,3,4,11,12,13).
\end{aligned}
\]
For each $k=1,2,3$, the noise matrices satisfy \((\Ev_k)_{i,j} \overset{iid}{\sim} \mathcal{N}(0,\sigma_k^2),\) with \(\sigma_1=\sigma_2=\sigma_3=1.\)

\subsection{Reconstruction of Korean mortality data} 

\begin{figure}[ht]
    \centering
    \includegraphics[width=1\linewidth]{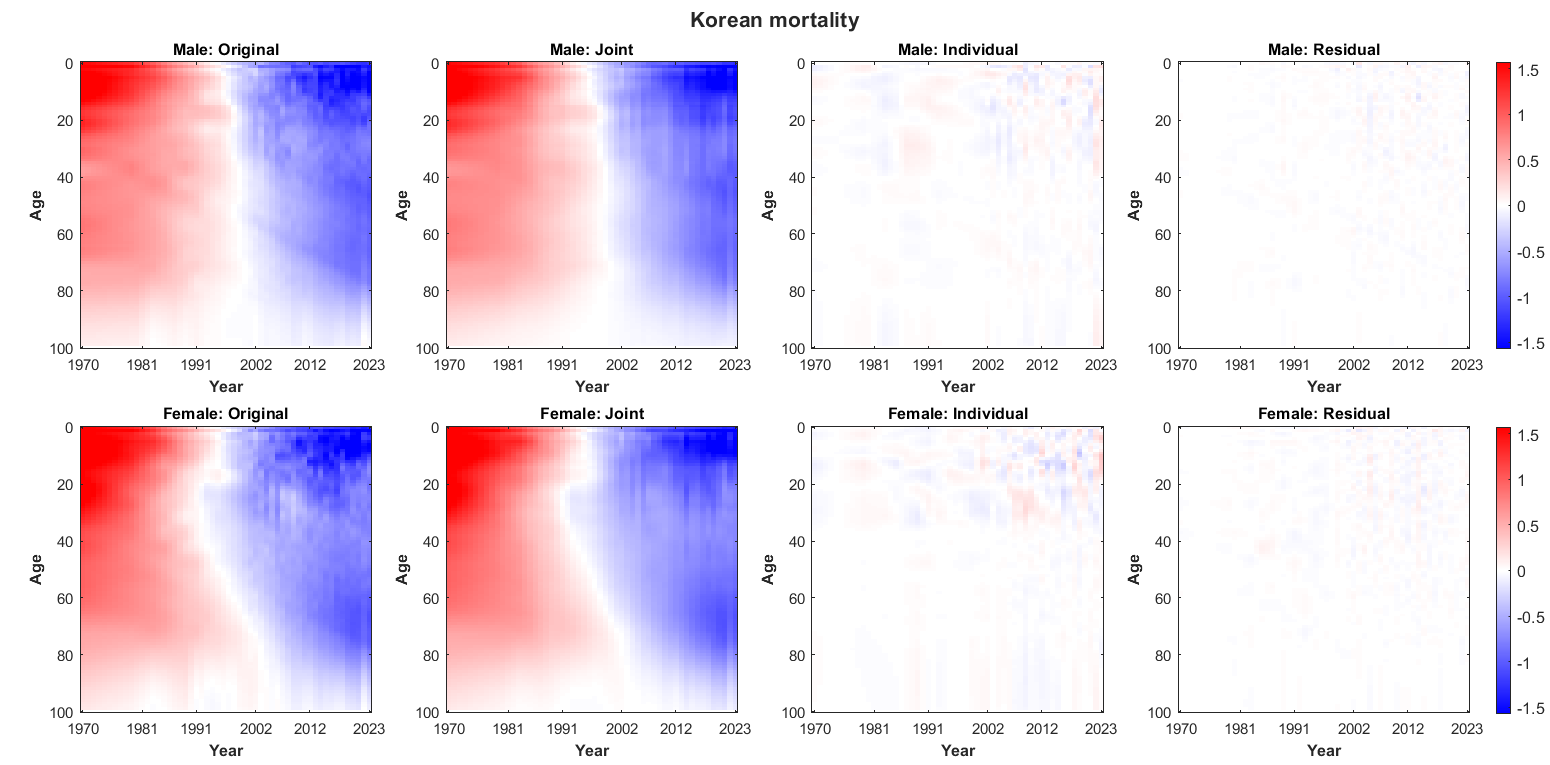}
    \caption{ 
    Korean mortality data (1969--2023): heatmap visualization of the observed matrices and the MSSAT decomposition.
    For each gender (male/female), the first column shows the row-centered $\log_{10}$ mortality matrix $\Xv_k$ (ages 0--100 by calendar year).
    The remaining columns display the reconstructed joint component $\hat{\Jv}_k$, individual component $\hat{\Iv}_k$, and residual $\hat{\Ev}_k=\Xv_k-\hat{\Jv}_k-\hat{\Iv}_k$.
    Color scales are indicated by the bars below each panel.
    }
    \label{fig:kore_mort_matrices}
\end{figure}

As referenced in Section 6 of the main article, we provide Figure~\ref{fig:kore_mort_matrices}, which visualizes the MSSAT reconstruction of the Korean mortality data using age--year heatmaps.
MSSAT selects a joint rank of $5$, which captures the dominant systematic variation and exhibits a highly coherent temporal pattern shared by both genders.
The individual components, with rank $6$ for each gender, are comparatively weaker in magnitude but display structured departures that are distinct from the residual noise, indicating remaining gender-specific signals after removing the shared component.
Finally, the residual matrices are largely featureless relative to the joint and individual layers, suggesting that the MSSAT decomposition accounts for most of the non-noise structure in each block.

\section{Technical details and proofs}\label{app:proof}

\subsection{Proof of Lemma~\xdocref{4}}
Proofs of Lemma~\xdocref{4} (i) and (ii) closely follow the arguments in \citet{ke2023estimation}. 
However, there is an important distinction in the underlying structure. 
While \citet{ke2023estimation}  assumes a spiked covariance model, our analysis is conducted under a noise-additive model.
\begin{proof}[Proof of Lemma~\xdocref{4} (i)]

    Let $\Wv_k = \Ev_k\Ev_k^\top/n$, $\Zv_k = \Xv_k\Xv_k^\top/n$ and $\ell_k(\beta) = [m_{k,\mathrm{low}}(\beta),\,m_{k,\mathrm{up}}(\beta)]$. Then
    \begin{align*}
        \Zv_k &= \frac{1}{n} \Xv_k\Xv_k^\top\\
        &= \frac{1}{n}\Ev_k\Ev_k^\top + \frac{1}{n}\left( \Av_k\Av_k^\top + \Av_k\Ev_k^\top + \Ev_k\Av_k^\top \right)\\
        &= \Wv_k + \frac{1}{n}\left( \Av_k\Av_k^\top + \Av_k\Ev_k^\top + \Ev_k\Av_k^\top \right),
    \end{align*}
    which shows that $\Zv_k$ is a perturbation of $\Wv_k$ by a matrix of rank at most $3r_k$.
    For large enough $n$, $\lambda_{l+3r_k}(\Wv_k)\leq \lambda_l\left(\Zv_k\right) \leq \lambda_{l-3r_k}(\Wv_k)$ for all $i \in I_k(\beta)$ by interlacing property.
    Meanwhile, by \citet{ke2023estimation}, $\sup_{l\in \ell_k(\beta)}| \lambda_l(\Wv_k) - \sigma_k^2 q_{k,l} | = O_\prec(n^{-1})$ holds.
    Those order in probability remains after replacing $\ell_k(\beta)$ by $\ell_k(\beta)^\circ = [\beta m_k - 3r_k,\, \beta m_k + 3r_k]$, as long as $\ell_k(\beta)^\circ$ lies in the bulk interior of the MP spectrum.
    Moreover, since the MP quantile map is strictly positive and smooth on the bulk interval $[\beta, 1-\beta]$, it is Lipschitz with constant $C_{k,\beta} >0$ satisfying $| q_{k,l\pm 3r_k} - q_{k,l}| \leq C_{k,\beta}3r_k/m_k=O(n^{-1})$ for all $l \in \ell_k(\beta)$. To put it together, 
    \begin{align*}
        0 &\leq \lambda_l\left(\widehat{\Sigma}_{\Sv,k}\right) - \lambda_{l+3r_k}\left(\Wv_k\right) \\
        &= \left( \lambda_l\left(\widehat{\Sigma}_{\Sv,k}\right) - \sigma_k^2 q_{k,l} \right) + \sigma_k^2\left( q_{k,l} - q_{k,l+3r_k} \right) + \left( \sigma_k^2 q_{k,l+3r_k} -  \lambda_{l+3r_k}(\Wv_k) \right), \\
        0 &\geq \lambda_l\left(\widehat{\Sigma}_{\Sv,k}\right) - \lambda_{l-3r_k}\left(\Wv_k\right) \\
        &= \left( \lambda_l\left(\widehat{\Sigma}_{\Sv,k}\right) - \sigma_k^2 q_{k,l} \right) + \sigma_k^2\left( q_{k,l} - q_{k,l-3r_k} \right) + \left( \sigma_k^2 q_{k,l-3r_k} -  \lambda_{l-3r_k}(\Wv_k) \right).
    \end{align*}
    As a result, by taking $\sup_{l \in \ell_k(\beta)}$, $\sup_{l\in \ell_k(\beta)} \left|\lambda_l \left( \Zv_k \right) - \sigma_k^2 q_{k,l}  \right| = O_\prec(n^{-1})$ holds.
    Finally, we have
    \begin{align*}
        \left| \hat{\sigma}_k^2 - \sigma_k^2\right| &= \left| \frac{\sum_{l\in \ell_k(\beta)}q_{k,l} \left( \lambda_l \left( \Zv_k\right) - \sigma_k^2 q_{k,l}\right) }{ \sum_{l\in \ell_k(\beta)} q_{k,l}^2 } \right| \\
        &\leq \sup_{l\in \ell_k(\beta)}\left|\left( \lambda_l \left( \Zv_k \right) - \sigma_k^2 q_{k,l}\right)\right| \frac{\sum_{l\in \ell_k(\beta)}q_{k,l} }{ \sum_{l\in \ell_k(\beta)} q_{k,l}^2 } = O_\prec\left(\frac{1}{n}\right),
    \end{align*}
    which proves Lemma~\xdocref{4} (i).
\end{proof}

\begin{proof}[Proof of Lemma~\xdocref{4} (ii)]

    Let $T_n = \hat{\sigma}_k^2\left[ (1+\sqrt{y_k})^2 + \operatorname{TW}_{1-\gamma_n} n^{-2/3} y_k^{-1/6} (1+\sqrt{y_k})^{3/4} \right]$ be the threshold in \xdocref{(10)}.
    By \citep{ding2020high}, for small $\epsilon_1>0$ and sufficiently large $n$, with probability $1-o(1)$,
    \begin{align*}
        \left|\frac{\sigma_{r_k}(\Xv_k)}{\sqrt{n}\sigma_k} - p_k(\lambda_{(k),r_k})\right|
        \leq C_1 n^{-1/2+\epsilon_1},
    \end{align*}
    for some constant $C_1$, where 
    \[
    p_k(s)=(1+s^2)\left(1+\frac{y_k}{s^2}\right).
    \]
    Meanwhile, $p_k(s)$ is monotone increasing for $s>\sqrt{y_k}$. Since $\lambda_{(k),r_k}^2>\sqrt{y_k}+\delta^2$ by assumption, we can write
    \begin{align*}
        p_k(\lambda_{(k),r_k})
        &= \left(1+\lambda_{(k),r_k}^2\right)\left(1+\frac{y_k}{\lambda_{(k),r_k}^2}\right) \\
        &= (1+\sqrt{y_k}) + \left( \lambda_{(k),r_k}^2 +\frac{y_k}{\lambda_{(k),r_k}^2}+ y_k - \sqrt{y_k} \right) \\
        % &\geq (1+\sqrt{y_k})^2 + \frac{\delta^2(2y_k^{1/4}+\delta)^2}{\lambda_{(k),r_k}^2}
        &\geq (1+\sqrt{y_k}) + \left( \frac{\delta^2(2\sqrt{y_k}+\delta)^2}{\lambda_{(k),r_k}^2} \right). \\
        &\geq (1+\sqrt{y_k}) + 4\frac{\delta^2 y_k}{C^2},
    \end{align*}
    where the last inequality follows from the assumption that $\lambda_{(k),r_k}\leq C$.

    To put it together, for large enough $n$, with probability $1-o(1)$, 
    \begin{align*}
        \frac{\sigma_{r_k}^2(\Xv_k)}{{n}} \geq \sigma_k^2 \left\{(1+\sqrt{y_k})^2 + \frac{\delta^4 y_k^2}{C^4}\right\},
    \end{align*}
    and by Lemma~\xdocref{4} (i), $|\hat{\sigma}_k^2 - \sigma_k^2| = O_\prec(1/n)$, we have 
    \begin{align*}
        \frac{\sigma_{r_k}^2(\Xv_k)}{{n}} \geq \hat{\sigma}^2_k \left\{ (1+\sqrt{y_k})^2 + \frac{\delta^4 y_k^2}{C^4}\right\}.
    \end{align*}

    Since $\operatorname{TW}_{1-\gamma_n} n^{-2/3} \to 0$ as $n\to \infty$, for large enough $n$, $T_n \leq \hat{\sigma}^2_k \left\{ (1+\sqrt{y_k})^2 + \frac{\delta^4 y_k^2}{C^4}\right\}$ which implies $\Pb(\sigma_{r_k}^2(\Xv_k)/n \geq T_n ) = 1-o(1)$.
 
    Also, by \citet{ding2020high}, for given $\epsilon>0$ and sufficiently large $n$, with probability $1-o(1)$,
    \begin{align*}
        \left|\frac{\sigma_{r_k+1}(\Xv_k)}{\sqrt{n}\sigma_k} - (1+\sqrt{y_k})\right|
        \leq n^{-2/3+\epsilon},
    \end{align*}
    so that by Lemma~\xdocref{4} (i), we have for some constant $C_2$,
    \begin{align*}
        \frac{\sigma_{r_k+1}(\Xv_k)}{n} \leq \hat{\sigma}_k^2\left\{ (1+\sqrt{y_k})^2 + C_2 n^{-2/3+\epsilon} \right\}.
    \end{align*}
    Since $\operatorname{TW}_{1-\gamma_n} n^{-\epsilon} \to \infty$, for large enough $n$, $T_n \geq \hat{\sigma}_k^2\left\{ (1+\sqrt{y_k})^2 + C_2 n^{-2/3+\epsilon} \right\}$ which implies $\Pb(\sigma_{r_k+1}^2(\Xv_k)/n \leq T_n ) = 1-o(1)$.    
\end{proof}

\begin{proof}[Proof of Lemma~\xdocref{4} (iii)]
    By \citet{benaych2012singular} and \citet{bao2021singular}, for all $l \in [r_k]$ and $k \in [K]$,
    \[ \left| {\lambda}_{(k),l} -  \frac{\hat{d}_{(k),l}}{\sqrt{n}\sigma_k} \right| = O_\prec\left(\frac{1}{\sqrt{n}}\right).\]
    Combining with the result of Lemma~\xdocref{4} (i), $|\hat{\lambda}_{(k),l} - \lambda_{(k),l}| = O_\prec(n^{-1/2})$ holds.    
\end{proof}

\subsection{Proof of Theorems~\xdocref{1} and \xdocref{2}}\label{sup:pf.thm}

Theorem~\xdocref{1} can be obtained as a special case of Theorem~\xdocref{2} by restricting to $K=2$.
Hence, we prove only Theorem~\xdocref{2}.
Before proving Theorem~\xdocref{2}, we present several lemmas and a corollary that will be used in its proof.

\begin{lem}\label{lem:unif.sph}

    Let $\left\{X_k\right\}_{k=1}^K$ be the set of random vectors each independently and uniformly distributed on the sphere $S^{n_k -1}$ where $n_k = n - r_k$ for some constants $0\leq r_1,\dots , r_K < n$. 
    For some constant $r$ such that $\max\left\{ r_1\dots, r_K \right\}\leq r < n $, set $n_0 = n-r$.
    Consider a random vector $Z_n := \sum_{k=1}^K w_k \pi_{n_0}(X_k)$ for some constants $w_1,\dots, w_K$.
    Here, with some abuse of notation, $\pi_{n_0} = \pi_{n_0}^{k}:\Rb^{n_k} \to \Rb^{n_0}$ is the natural projection that takes the first $n_0$ elements of input vector. Then the following holds:
    \begin{align*}
        \sqrt{n}\left(\|Z_n\|_2^2 - \sum_{k=1}^K w_k^2\right) \overset{d}{\to} \Nc\left(0,2\left\{\sum_{k=1}^K w_k^2 \right\}^2 - 2\sum_{k=1}^K w_k^4 \right) \quad\textnormal{as } n\to \infty
    \end{align*}

\end{lem}

\begin{proof}
    Each $X_k$ follows uniform on the sphere $S^{n_k-1}$, then has same distribution with $Y_k/\|Y_k\|_2$ where $Y_k \overset{ind}{\sim} \Nc_{n_k}(0,I_{n_k})$. Let $[\cdot]_i$ be the $i$-th element of vector. Since the $i$-th element $[X_k]_i \overset{d}{=}[Y_k/\|Y_k\|_2]_i = [Y_k]_i/\sqrt{n} + O_p(1/n)$, ${\pi_{n_0}( Y_k)}/{\|Y_k\|_2}$ can be decomposed as:
    \begin{align*}
        \pi_{n_0}(X_k) \overset{d}{=} \frac{\pi_{n_0}( Y_k)}{\|Y_k\|_2}  =  \frac{1}{\sqrt{n}}\pi_{n_0}(Y_k) + R_k
    \end{align*}
    where $R_k := \pi_{n_0}( Y_k)(1/\|Y_k\|_2 - 1/\sqrt{n})$ which has $O_p(1/n)$ element and $0\leq \|R_k\|_2 \leq \|Y_k\|_2(1/\|Y_k\|_2 - 1/\sqrt{n}) = O_p(1/\sqrt{n})$. Note that $Y_k$ and $R_l$ are independent for $k \neq l$. Meanwhile,
    \begin{align*}
        \|Z_n\|_2^2 = \left\| \sum_{k=1}^K w_k\pi_{n_0}(X_k) \right\|_2^2 &= \left\{ \sum_{k=1}^K w_k^2 \| \pi_{n_0}(X_k)\|^2 \right\} + \sum_{k\neq l}w_kw_l\langle \pi_{n_0}(X_k), \pi_{n_0}(X_l)\rangle.
    \end{align*}
    For the first part, $\| \pi_{n_0}(X_k)\|^2 =  1 - \sum_{i=n_0+1}^{n_k} [X_k]_i^2 = 1+O_p(1/n)$. For the second part, $k \neq l$ then, 
    \begin{align*}
        \langle \pi_{n_0}(X_k), \pi_{n_0}(X_l)\rangle &\overset{d}{=} \frac{1}{n}\langle \pi_{n_0}(Y_k), \pi_{n_0}(Y_l)\rangle + \frac{1}{\sqrt{n}}\langle \pi_{n_0}(Y_k), R_l \rangle + \frac{1}{\sqrt{n}}\langle R_k,\pi_{n_0}(Y_l) \rangle + \langle R_k,R_l\rangle \\
        &= \frac{1}{n}\langle \pi_{n_0}(Y_k), \pi_{n_0}(Y_l)\rangle + O_p\left(\frac{1}{n}\right).
    \end{align*}
    Here we use the fact that $\langle \pi_{n_0}(Y_k), R_l/\|R_l\| \rangle \sim \Nc(0,1)$ and $|\langle R_k, R_l \rangle| \leq \|R_k\|_2\|R_l\|_2 = O_p(1/n)$. 
    To put it together,
    \[\|Z_n\|_2^2 \overset{d}{=} \sum_{k=1}^Kw_k^2 + \frac{1}{n}\sum_{k\neq l} w_k w_l \langle \pi_{n_0}(Y_k), \pi_{n_0}(Y_l) \rangle + O_p\left(\frac{1}{n}\right).\]
    Finally, the following convergence in distribution, obtained via the central limit theorem, implies the desired result:
    \begin{align*}
        \frac{1}{\sqrt{n}}\sum_{k\neq l} w_k w_l \langle \pi_{n_0}(Y_k), \pi_{n_0}(Y_l) \rangle &= \frac{1}{\sqrt{n}}\sum_{k\neq l} w_k w_l \left\{\sum_{i=1}^{n_0} [Y_k]_i[Y_l]_i \right\} \\
        &= \sqrt{n}\left[ \frac{1}{n}\sum_{i=1}^{n_0} \sum_{k\neq l} w_k w_l [Y_k]_i[Y_l]_i\right] \\
        &\overset{d}{\to} \Nc\left(0, 2\left\{\sum_{k=1}^K w_k^2 \right\}^2 - 2\sum_{k=1}^K w_k^4 \right) \quad\textnormal{as } n\to \infty,
    \end{align*}
    since $\Eb[\sum_{k\neq l} w_k w_l [Y_k]_i[Y_l]_i] = 0$ and $\Eb[\sum_{k\neq l} w_k w_l [Y_k]_i[Y_l]_i]^2 = \sum_{k\neq l} 2 w_k^2 w_l^2$.

\end{proof}

Allowing the weights to be random but consistent leads to the following immediate extension.

\begin{cor}\label{cor:unif.sph}
    Let $\left\{ w_k \right\}_{k=1}^{K}$ be random weights such that $w_k \overset{p}{\to} s_k$ for some constant $\left\{s_k\right\}_{k=1}^K$. Then for $Z_n$ defined in Lemma~\ref{lem:unif.sph}, the following holds.
    \begin{align*}
        \sqrt{n}\left( \|Z_n\|_2^2 - \sum_{k=1}^Kw^2_k \right) \overset{d}{\to} \Nc\left(0,2\left\{\sum_{k=1}^K s_k^2 \right\}^2 - 2\sum_{k=1}^K s_k^4 \right) \quad\textnormal{as } n\to \infty
    \end{align*}
\end{cor}

\begin{proof}
    Considering $\langle \pi_{n_0}(X_k), \pi_{n_0}(X_l) \rangle = O_p(1/\sqrt{n})$ for $k\neq l$,
    \begin{align*}
        \sqrt{n}\left\{ \|Z_n\|_2^2 -\sum_{k=1}^Kw_k^2 \right\} &= \sqrt{n}\sum_{k\neq l} w_k w_l \langle \pi_{n_0}(X_k), \pi_{n_0}(X_l) \rangle + O_p\left(\frac{1}{\sqrt{n}}\right) \\
        & = \sqrt{n}\sum_{k\neq l} s_k s_l \langle \pi_{n_0}(X_k), \pi_{n_0}(X_l) \rangle + o_p(1)
    \end{align*}
    where the second equality follows from $w_k \overset{p}{\to} s_k$ and the boundedness of the inner products. 
    The remaining argument is identical to that in the proof of Lemma~\ref{lem:unif.sph}, yielding the stated Gaussian limit.
\end{proof}

\begin{lem}\label{lem:rot.inv}
    Consider the noise additive model $\Xv = \Av + \Ev$, where $\Av = \Uv_{(p\times r) }\Dv_{(r \times r) }\Vv^\top_{(r \times n)}$ is a rank-$r$ fixed matrix and $\Ev$ has distribution that is right-orthogonally invariant i.e. $\Lc(\Ev) = \Lc(\Ev \Rv)$ for all $\Rv \in \Oc(n)$. Let $\Vv_\perp \in \Oc(n,n-r)$ be any orthonormal matrix orthogonal to $\Vv$, that is, $\Vv^\top \Vv_\perp = \boldsymbol{0}$. Then, for each $l = 1,\dots,r$, the normalized projection
    \[
        \frac{\Vv_\perp^\top \vv_l(\Xv)}{\|\Vv_\perp^\top \vv_l(\Xv)\|}
    \]
    is uniformly distributed on the unit sphere $\Sc^{n-r-1}$ and independent of $\Vv^\top \vv_l(\Xv)$.
\end{lem}

\begin{proof}
The argument parallels the proof of Theorem~6 in \citet{paul2007asymptotics}. 
Let $\Rv = \Vv\Vv^\top + \Vv_\perp \Rv_{n-r} \Vv_\perp^\top$ for an arbitrary $\Rv_{n-r}\in \Oc(n-r)$.  
Then, since $\Rv\Vv = \Vv$ and the noise is right-rotation invariant,
\[
\Lc(\Xv \Rv^\top)
= \Lc(\Av \Rv^\top + \Ev \Rv^\top)
= \Lc(\Av + \Ev)
= \Lc(\Xv).
\]
From the singular vector identity $\vv_l(\Xv \Rv^\top) = \Rv \vv_l(\Xv)$, it follows that
$\Vv^\top \vv_l(\Xv \Rv^\top) = \Vv^\top \Rv \vv_l(\Xv) = \Vv^\top \vv_l(\Xv)$.  
Hence, conditioning on $\Vv^\top \vv_l(\Xv)$,
\begin{equation}\label{eq:rot.inv}
\begin{aligned}
\Lc(\Vv_\perp^\top \vv_l(\Xv)\mid \Vv^\top \vv_l(\Xv))
&= \Lc(\Vv_\perp^\top \vv_l(\Xv \Rv^\top)\mid \Vv^\top \vv_l(\Xv \Rv^\top)) \\
&= \Lc(\Vv_\perp^\top \Rv \vv_l(\Xv)\mid \Vv^\top \vv_l(\Xv)) \\
&= \Lc(\Rv_{n-r} \Vv_\perp^\top \vv_l(\Xv)\mid \Vv^\top \vv_l(\Xv)).
\end{aligned}
\end{equation}
Because $\Rv_{n-r}$ is arbitrary in $\Oc(n-r)$, the conditional law in~\eqref{eq:rot.inv}
must be invariant under all right-orthogonal transformations in $\Rb^{n-r}$.  
Therefore, 
$\Vv_\perp^\top \vv_l(\Xv)/\|\Vv_\perp^\top \vv_l(\Xv)\| \mid \Vv^\top \vv_l(\Xv)$ 
is uniformly distributed on the unit sphere $\Sc^{n-r-1}$ and is independent of $\Vv^\top \vv_l(\Xv)$.
\end{proof}

Lastly, we introduce the asymptotic distribution of $\angle(\hat{\vv}_{\Av_k,\ell_k}, \vv_{\Av_k,\ell_k})$, the angle between the true and estimated singular vectors, established by \citet{bao2021singular}.
For clarity, we restate a simplified version of their main theorem below.

\begin{lem}(Asymptotic distribution of angles, \citep{bao2021singular})\label{thm:asym.dist.angles.Bao}
    
    Under Assumption~\xdocref{1}, for all $i \in [r_k]$,
    \begin{align}\label{eq:clt.bao}
        \sqrt{\frac{n}{\Vc^\circ_{y_k}(\lambda_{(k),i})}}\left( \cos^2 \angle(\hat{\vv}_{(k),i}, \vv_{(k),i})  - m^2_{y_k} (\lambda_{(k),i}) \right) \overset{d}{\to} \Nc\left( 0, 1 \right),
    \end{align}
    where %the $\Vc_{y_1,y_2}(\lambda_1,\lambda_2)$ is defined as follows:
    \begin{align*}
        &m_y(s) := \sqrt{\frac{s^4-y}{s^2(s^2+1)}}, \quad\theta_y(s) :=\frac{s^4+2 y s^2+y}{s^3\left(s^2+1\right)^2}, \quad \psi_y(s):=\frac{s^6-3 y s^2-2 y}{s^3\left(s^2+1\right)^2}, \\
        &\mathcal{V}_y(s):=  \frac{s^2(s^2+1)}{2(s^4-y)^2} \left(2 y(y+1) \theta_y(s)^2-\frac{y(y-1)(5 y+1)}{s\left(s^2+1\right)^2} \theta_y(s)\right. \\
        &\quad\quad\quad\quad\left.+\frac{\left(s^4+y\right)\left(s^2+y\right)^2}{s^3\left(s^2+1\right)^2} \psi_y(s) +\frac{2 y^2(y-1)^2}{s^2\left(s^2+1\right)^4}\right) + \frac{\theta^2_y(s)}{m_y^2(s)},\\
        & \Vc_y^\circ(s):= 4\Vc_y(s)m^2_y(s).
    \end{align*}
    \eqref{eq:clt.bao} can be rewritten as 
    \begin{align*}
        \sqrt{\frac{n}{\Vc_{y_k}(\lambda_{(k),i})}}\left(\left| \langle\hat{\vv}_{(k),i}, \vv_{(k),i}\rangle\right|  - {m_{y_k} (\lambda_{(k),i})}  \right) \overset{d}{\to} \Nc\left( 0, 1 \right).
    \end{align*}
    Furthermore, $\cos^2 \angle(\hat{\vv}_{(k),i}, \vv_{(k),j}) = O_\prec(n^{-1})$ for all $ i\neq j \in [r_k]$.
\end{lem}

\begin{proof}[Proof of Theorem~\xdocref{2}]
    Write $\vv_{\textsc{j}} = \vv_{(1),\ell_1}=\cdots = \vv_{(K),\ell_K}$.
    Let $S:= \cup_{k=1}^K \operatorname{span}(\Vv_{(k)}) \setminus \operatorname{span}(\vv_{\textsc{j}})$ be the subspace with dimension $r_s = \sum_{k=1}^K (r_k-r_{\textsc{j}}) +(r_{\textsc{j}}-1)$. Consider $\left\{ \vv_{\textsc{j}}, \vv_{s_1}, \dots, \vv_{s_{r_s}}, \vv_{e_1},\dots, \vv_{e_{r_e}} \right\}$ a basis of $\Rb^n$ where $\left\{ \vv_{s_1}, \dots, \vv_{s_{r_s}} \right\}$ span $S$. Note that $r_e = n-\sum_{k=1}^K(r_k - r_{\textsc{j}}) -r_{\textsc{j}}$ and each vector in $\left\{ \vv_{e_1},\dots, \vv_{e_{r_e}} \right\}$ is orthogonal to $\Vv_{(k)}$ for all $k=1,\dots, K$. By Parseval's identity, we can write:
    \begin{align*}
        \hat{\vv}_{(k),\ell_k} = \langle \hat{\vv}_{(k),\ell_k}, \vv_{\textsc{j}} \rangle\vv_{\textsc{j}} + \sum_{i=1}^{r_s} \langle\hat{\vv}_{(k),\ell_k}, \vv_{s_i}\rangle \vv_{s_i} + \sum_{j=1}^{r_e} \langle\hat{\vv}_{(k),\ell_k}, \vv_{e_j} \rangle \vv_{e_j}
    \end{align*}
    and 
    \begin{align}\label{eq:parseval}
        \|\sum_{k=1}^K \hat{\vv}_{(k),\ell_k}\|_2^2 &= \left\{ \sum_{k=1}^K \langle\hat{\vv}_{(k),\ell_k},\vv_{\textsc{j}} \rangle \right\}^2 + \sum_{i=1}^{r_s} \left\{\sum_{k=1}^K \langle\hat{\vv}_{(k),\ell_k}, \vv_{s_i}\rangle \right\}^2 + \sum_{j=1}^{r_e} \left\{ \sum_{k=1}^K\langle\hat{\vv}_{(k),\ell_k}, \vv_{e_j}\rangle \right\}^2.
    \end{align}
        
    For the second term in (\ref{eq:parseval}), according to \cite{bao2021singular}, we have $|\langle \hat{\vv}_{(k),\ell_k},\vv_{(k),l} \rangle| = O_\prec(1/\sqrt{n})$ for all $l \neq \ell_k$. Moreover, by Lemma~\ref{lem:rot.inv}, $|\langle \hat{\vv}_{(k),\ell_k},\vv \rangle| = O_p(1/\sqrt{n})$ for any $\vv$ orthogonal to $\Vv_{(k)}$. 
    Hence, since each $\vv_{s_i}$ is a linear combination of finitely many vectors $\vv_{(k),l}$ and $\vv$ orthogonal to $\Vv_{(k)}$, it follows that $\langle\hat{\vv}_{(k),\ell_k}, \vv_{s_i}\rangle = O_\prec(1/\sqrt{n})$ for all $i =1,\dots, r_s$. Then,
    \begin{align*}
        \sum_{i=1}^{r_s} \left\{\sum_{k=1}^K \langle\hat{\vv}_{(k),\ell_k}, \vv_{s_i}\rangle \right\}^2 \leq \sum_{i=1}^{r_s} \left\{\sum_{k=1}^K K \langle\hat{\vv}_{(k),\ell_k}, \vv_{s_i}\rangle^2 \right\} = O_\prec\left(\frac{1}{n}\right).
    \end{align*}

    For the last term of (\ref{eq:parseval}), let $\Vv_{res}:= [\vv_{e_1}|\cdots|\vv_{e_{r_e}}]$, the orthonormal matrix consisting of columns $\left\{\vv_{e_j}\right\}_{j=1}^{r_e}$. Then, the last term of (\ref{eq:parseval}) can be represented as:
    \begin{align*}
        \sum_{j=1}^{r_e} \left\{ \sum_{k=1}^K\langle\hat{\vv}_{(k),\ell_k}, \vv_{e_j}\rangle \right\}^2 = \left\| \sum_{k=1}^K \Vv_{res}^\top \hat{\vv}_{(k),\ell_k} \right\|_2^2.
    \end{align*}
    From Lemma~\ref{lem:rot.inv}, $\Vv_{res}^\top \hat{\vv}_{(k),\ell_k}\left/\sqrt{1-\| \Vv_{(k)}^\top \hat{\vv}_{(k),\ell_k} \|^2}\right.$ has same distribution with $\pi_{r_e}(S_k)$ where $S_k$ is random variables uniformly distributed on the unit sphere $\Sc^{r_e -1 }$ and independent to $\Vv_{(k)}^\top \hat{\vv}_{(k),\ell_k}$. From Corollary~\ref{cor:unif.sph}, with the fact that, $ \sqrt{1-\| \Vv_{(k)}^\top \hat{\vv}_{(k),\ell_k} \|^2} \overset{p}{\to} \sqrt{1-m^2_{y_k}(\lambda_{(k),\ell_k})}$ as $n \to \infty$,
    \begin{align*}
        \sqrt{n}&\left( \sum_{j=1}^{r_e} \left\{ \sum_{k=1}^K\langle\hat{\vv}_{(k),\ell_k}, \vv_{e_j}\rangle \right\}^2 \right. - \left. \sum_{k=1}^K\left(1-\| \Vv_{(k)}^\top \hat{\vv}_{(k),\ell_k} \|^2\right) \right) \\
        &\overset{d}{\to} \Nc\left( 0, 2\left\{ \sum_{k=1}^K \left(1- m^2_{y_k}(\lambda_{(k),\ell_k}) \right) \right\}^2 - 2 \sum_{k=1}^K \left\{1-m^2_{y_k}(\lambda_{(k),\ell_k})\right\} ^2 \right).
    \end{align*}

    For the first term of (\ref{eq:parseval}), \cite{bao2021singular} find asymptotic distribution of each $\langle\hat{\vv}_{(k),\ell_k},\vv_{\textsc{j}} \rangle$ which is stated in Lemma~\ref{thm:asym.dist.angles.Bao}. Since the data blocks $\left\{\Xv_k \right\}_{k=1}^K$ are all mutually independent, $\left\{\langle\hat{\vv}_{(k),\ell_k},\vv_{\textsc{j}} \rangle \right\}_{k=1}^K$ are also mutually independent. From these, we have:
    \begin{align*}
        &\sqrt{n}\left(\left\{ \sum_{1\leq k\neq l \leq K} \langle\hat{\vv}_{(k),\ell_k},\vv_{\textsc{j}} \rangle\langle\hat{\vv}_{(l),\ell_l},\vv_{\textsc{j}} \rangle \right\} - \sum_{1\leq k\neq l \leq K}m_{y_k}(\lambda_{(k),\ell_k})m_{y_l}(\lambda_{(l),\ell_l})\right) \\
        &=\sqrt{n}\left(\left\{ \sum_{k=1}^K \langle\hat{\vv}_{(k),\ell_k},\vv_{\textsc{j}} \rangle \right\}^2\right.+ \left.\sum_{k=1}^K\left( 1- \langle \hat{\vv}_{(k),\ell_k},\vv_{\textsc{j}} \rangle^2\right)\right. \\
        &\left. \quad\quad\quad\quad-\left\{\sum_{k=1}^K {m_{y_k}(\lambda_{(k),\ell_k})} \right\}^2 - \sum_{k=1}^K\left( 1-m^2_{y_k} (\lambda_{(k),\ell_k}) \right)  \right) \\
        &=\sqrt{n}\left( \sum_{k=1}^K\left(\langle\hat{\vv}_{(k),\ell_k},\vv_{\textsc{j}} \rangle - {m_{y_k}(\lambda_{(k),\ell_k})}\right)\left\{2 \sum_{l\neq k} {m_{y_l}(\lambda_{(l),\ell_l})} \right\} + O_p\left(\frac{1}{n}\right) \right)\\
        &\overset{d}{\to} \Nc\left(0,4\sum_{k=1}^K {\Vc_{y_k}(\lambda_{(k),\ell_k})}\left\{ \sum_{l\neq k} {m_{y_l}(\lambda_{(l),\ell_l})} \right\}^2\right).
    \end{align*}

    To put it together with ${\| \Vv_{(k)}^\top \hat{\vv}_{(k),\ell_k} \|^2 -  \langle\hat{\vv}_{(k),\ell_k},\vv_{\textsc{j}} \rangle^2 } = O_\prec(1/n)$, we have:
    \begin{align*}
        &\sqrt{n}\left\{  \left\|\sum_{k=1}^K \hat{\vv}_{(k),\ell_k}\right\|_2^2 - \left(K + \sum_{k\neq l}{m_{y_k}(\lambda_{(k),\ell_k})m_{y_l}(\lambda_{(l),\ell_l})} \right) \right\}\\ 
        &\overset{d}{\to} \Nc\left(0,4\sum_{k=1}^K {\Vc_{y_k}(\lambda_{(k),\ell_k})}\left\{ \sum_{l\neq k} {m_{y_l}(\lambda_{(l),\ell_l})} \right\}^2 + 4\sum_{k < l}(1-m^2_{y_k}(\lambda_{(k),\ell_k}))(1-m^2_{y_l}(\lambda_{(l),\ell_l}))\right)
    \end{align*}
    Considering $\Vc_{y}(\lambda)$ and $m_y(\lambda)$ are positive and continuous with respect to $y$, the desired result holds.
\end{proof}

\subsection{Proof of Theorem~\xdocref{3}}\label{sec:power_one}
Before proving Theorem~\xdocref{3}, we present a lemma that explains the limit of the global alignment statistic under the alternative.

\begin{lem}[Limit of the global alignment statistic under the alternative]\label{lem:alter.limit}
Suppose the conditions in Assumption~\xdocref{1} hold for each $X_k$, $k=1,\dots,K$.
Under $H_1$,
\begin{align}\label{eq:alt.limit}
T^2_{\boldsymbol\ell} = \mu_{\yv}(\lambdav)-\Delta_\ell + O_p(n^{-1/2}),
\end{align}
where $\Delta_\ell := 2\sum_{k<l}{m_{y_k}(\lambda_{(k),\ell_k})\,m_{y_l}(\lambda_{(l),\ell_l})}\,\bigl(1-\cos\phi_{\ell_k,\ell_l}\bigr) \;>\; 0.$
\end{lem}

\begin{proof}
    Note $T^2_{\boldsymbol\ell} = \| \sum_{k=1}^K \hat{\vv}_{(k),\ell_k} \|_2^2 = K + \sum_{k \neq l } \langle\hat{\vv}_{(k),\ell_k}, \hat{\vv}_{(l),\ell_l} \rangle$. Focusing on $\langle\hat{\vv}_{(k),\ell_k}, \hat{\vv}_{(l),\ell_l} \rangle$ for fixed $k$ and $l$ such that $k \neq l$, we consider two basis of $\Rb^n$,
    \begin{align*}
        &\left\{ \vv_{(k),\ell_k}, \vv_{(k),s_1}, \dots, \vv_{(k),s_{r_s}}, \vv_{e_1}, \dots, \vv_{e_{r_e}} \right\} \; \textnormal{ and } \\
        &\left\{ \vv_{(l),\ell_l}, \vv_{(l),s_1}, \dots, \vv_{(l),s_{r_s}}, \vv_{e_1}, \dots, \vv_{e_{r_e}} \right\},
    \end{align*}
     where both $\left\{ \vv_{(k),\ell_k}, \vv_{(k),s_1}, \dots, \vv_{(k),s_{r_s}} \right\}$ and $\left\{ \vv_{(l),\ell_l}, \vv_{(l),s_1}, \dots, \vv_{(l),s_{r_s}} \right\}$ span the same subspace $ \operatorname{span}(\Vv_{(k)}) \cup \operatorname{span}(\Vv_{(l)})$. Note that $r_s = r_k+r_l-r_{\textsc{j}}-2$ which is finite. Analogous to the proof of Theorem~\xdocref{2}, we wirte:
    \begin{align*}
        \hat{\vv}_{(k),\ell_k} &= \langle \hat{\vv}_{(k),\ell_k}, {\vv}_{(k),\ell_k} \rangle {\vv}_{(k),\ell_k} + \sum_{i=1}^{r_s}\langle \hat{\vv}_{(k),\ell_k}, {\vv}_{(k),s_i} \rangle {\vv}_{(k),s_i} + \sum_{j=1}^{{r_e}}\langle \hat{\vv}_{(k),\ell_k}, {\vv}_{e_j} \rangle {\vv}_{e_j} \\
        \hat{\vv}_{(l),\ell_l} &= \langle \hat{\vv}_{(l),\ell_l}, {\vv}_{(l),\ell_l} \rangle {\vv}_{(l),\ell_l} + \sum_{i=1}^{r_s}\langle \hat{\vv}_{(l),\ell_l}, {\vv}_{(l),s_i} \rangle {\vv}_{(l),s_i} + 
        \sum_{j=1}^{{r_e}}\langle \hat{\vv}_{(l),\ell_l}, {\vv}_{e_j} \rangle {\vv}_{e_j}
    \end{align*}
    
    Considering $\langle \hat{\vv}_{(k),\ell_k}, {\vv}_{(k),s_i} \rangle = O_\prec(1/\sqrt{n})$ and $\langle \hat{\vv}_{(l),\ell_l}, {\vv}_{(l),s_i} \rangle = O_\prec(1/\sqrt{n})$ for $i=1,\dots, r_s$ and the result of Corollary~\ref{cor:unif.sph}, we have:
    \begin{align*}
        \langle \hat{\vv}_{(k),\ell_k}, \hat{\vv}_{(l),\ell_l}\rangle &= \langle \hat{\vv}_{(k),\ell_k}, {\vv}_{(k),\ell_k} \rangle \langle \hat{\vv}_{(l),\ell_l}, {\vv}_{(l),\ell_l} \rangle \langle {\vv}_{(k),\ell_k}, {\vv}_{(l),\ell_l} \rangle + O_p\left(\frac{1}{\sqrt{n}}\right)\\
        &= {m_{y_k}(\lambda_{(k),\ell_k}) m_{y_l}(\lambda_{(l),\ell_l})} \cos \phi_{kl} + O_p\left(\frac{1}{\sqrt{n}}\right)
    \end{align*}

    As a result, 
    \begin{align*}
        T^2_{\boldsymbol\ell} &= K + \sum_{k \neq l } \langle\hat{\vv}_{(k),\ell_k}, \hat{\vv}_{(l),\ell_l} \rangle\\
        &= K + \sum_{k\neq l}\sqrt{a_{y_k}(\lambda_{(k),\ell_k})\,a_{y_l}(\lambda_{(l),\ell_l})}\,\cos\phi_{kl} + O_p\left(\frac{1}{\sqrt{n}}\right),
    \end{align*}
    which proves the claim.
\end{proof}

\begin{proof}[Proof of Theorem~\xdocref{3}]

Since $\hat{\lambda}_{(k),\ell_k} = {\lambda}_{(k),\ell_k} + O_p(n^{-1/2})$, we have ${\Vc}^\circ_{\yv}(\widehat{\lambdav}) \overset{p}{\to} {\Vc}^\circ_{\yv}({\lambdav})$ and $\mu_{\yv}(\widehat{\lambdav}) = \mu_{\yv}({\lambdav}) + O_p(n^{-1/2})$. Furthermore, the supercritical condition ensures that $\sqrt{{\Vc}^\circ_{\yv}({\lambdav}) / {\Vc}^\circ_{\yv}(\widehat{\lambdav})} = 1+ o_p(1)$. Then, by Lemma~\ref{lem:alter.limit},
\begin{align*}
    \widehat{Z}_{\boldsymbol\ell} & = \sqrt{\frac{n}{{\Vc}^\circ_{\yv}(\widehat{\lambdav})} } (T^2 - \hat\mu_\yv(\widehat{\lambdav})) \\
    & = \sqrt{\frac{n}{{\Vc}^\circ_{\yv}(\widehat{\lambdav})} } \left(\mu_\yv(\lambdav) - \hat\mu_\yv(\widehat{\lambdav}) - \Delta_{\boldsymbol{\ell}} +O_p\left(\frac{1}{\sqrt{n}}\right)\right) \\
    &=-\sqrt{n}\Delta_{\boldsymbol{\ell}}\left\{{\Vc}^\circ_{\yv}({\lambdav})\right\}^{-1/2}\left(1+o_p(1)\right) + O_p(n^{-1/2})
\end{align*}
Since ${\Vc}^\circ_{\yv}({\lambdav})>0$, and by the assumption of the theorem, we have $\sqrt{n}\Delta_{\boldsymbol{\ell}} \to \infty$, thus for any $c \in \Real$, $\lim_{n,p_k\to\infty} \mathbb{P}( \widehat{Z}_{\boldsymbol\ell} \le c)  = 1$, as required. 
\end{proof}
 
\subsection{Proof of Theorem \xdocref{5}}

\begin{proof}

First, we show that $\langle \tilde{\vv}_{(k)}^\ast, \Vv_{(k)} \wv_{(k)}^\perp \rangle = O_P(n^{-1/2})$ for all unit vectors $\wv_{(k)}^\perp$ orthogonal to $\wv_{(k)}$. Moreover, for any $\Vv_{(k)}^\perp \in O(n,n-r_k)$ satisfying $\Vv_{(k)}^\top \Vv_{(k)}^\perp = \boldsymbol{0}$, the normalized projection
\[
\frac{(\Vv_{(k)}^\perp)^\top \tilde{\vv}_{(k)}^\ast}{\|(\Vv_{(k)}^\perp)^\top \tilde{\vv}_{(k)}^\ast\|_2}
\]
is uniformly distributed on the unit sphere $\Sc^{n-r_k-1}$ and independent of $\|\Vv_{(k)}^\top \tilde{\vv}_{(k)}^\ast\|_2$.
The remaining proofs of (a) and (b) then follow along the same lines as those of Lemma~\ref{lem:alter.limit} and Theorem~\xdocref{2}, respectively. 

For the first claim, recall that $\tilde{\vv}_{(k)}^\ast=\hat{\Vv}_{(k)}\wv_{(k)}^\circ$ with $\wv_{(k)}^\circ=\Dv_k^{-1}\wv_{(k)}/\|\Dv_k^{-1}\wv_{(k)}\|_2$. 
From Lemma~\ref{thm:asym.dist.angles.Bao}, $\hat{\Vv}_{(k)}^\top \Vv_{(k)} = \Dv_k + O_p(n^{-1/2})$ where the $O_p(n^{-1/2})$ term is understood entrywise.
Recalling that the supercritical condition in Assumption~\xdocref{1} guarantees that each diagonal element of $\Dv_k$ is uniformly bounded away from both zero and infinity, the following hold:
\begin{align*}
    \langle  \tilde{\vv}_{(k)}^\ast, \Vv_{(k)} \wv_{(k)}^\perp \rangle
    &= \langle  \hat{\Vv}_{(k)}\wv_{(k)}^\circ, \Vv_{(k)} \wv_{(k)}^\perp \rangle \\
    &= \frac{1}{\|\Dv_k^{-1}\wv_{(k)}\|_2}\langle  \hat{\Vv}_{(k)}\Dv_k^{-1}\wv_{(k)}, \Vv_{(k)} \wv_{(k)}^\perp \rangle \\
    &= \frac{1}{\|\Dv_k^{-1}\wv_{(k)}\|_2} \wv_{(k)}^\top \Dv_k^{-1} \hat{\Vv}_{(k)}^\top \Vv_{(k)} \wv_{(k)}^\perp  \\
    &= \frac{1}{\|\Dv_k^{-1}\wv_{(k)}\|_2} \left\{\wv_{(k)}^\top \wv_{(k)}^\perp + O_p\left(n^{-1/2}\right)\right\} \\
    &= O_p\left( n^{-1/2}\right).
\end{align*}

For the second claim, the argument in Lemma~\ref{lem:rot.inv} remain valid after replacing $\vv_l(\cdot)$ by $\vv_{\wv}(\cdot):\Rb^{p\times n}\to\Sc^{n-1}$ defined by $\vv_{\wv}(\Xv)=\hat{\Vv}_r(\Xv)\wv$, where $\hat{\Vv}_r(\Xv)$ denotes the rank-$r$ right singular vector matrix of $\Xv$ and $\wv\in\Sc^{r-1}$ is fixed.
In particular, conditional on the signal, the noise is rotationally invariant in the orthogonal complement of $\operatorname{span}(\Vv_k)$, which implies that the direction of $\Vv_{k,\perp}^\top \tilde{\vv}_{(k),s}$ after normalization is uniform on $\Sc^{n-r_k-1}$ and independent of its length.
    
\end{proof}

\subsection{Motivation and approximation of the sequential alignment test}\label{sec:int.seq.test}

The purpose of the following discussion is to clarify the approximation used in \xdocref{(18)}. 
The leading term in the expansion of $\langle \vv_{\textsc{j}},\tilde{\vv}_{(k)}^\ast\rangle$ is the diagonal contribution
\[
    \sum_{i=1}^{r_k} w_{(k),i}w_{(k),i}^{\circ}
    \langle \vv_{(k),i},\hat{\vv}_{(k),i}\rangle ,
\]
whose mean is directly determined by the asymptotic singular vector alignment $m_{y_k}(\lambda_{(k),i})$. 
This yields the plug-in form of $\hat m_{(k)}$ in \xdocref{(18)}. 
The remaining cross-product terms $\langle \vv_{(k),i},\hat{\vv}_{(k),j}\rangle$ for $i\neq j$ are asymptotically smaller and do not affect the leading-order mean, but their joint contribution to the exact variance is more delicate, especially because the empirical weights used in the actual procedure are themselves random. 
For this reason, we use the oracle construction to justify the mean exactly at the required order, and then use a working variance approximation obtained by retaining only the diagonal fluctuation terms. 
The derivation below formalizes this approximation and explains how it leads to the empirical variance formula in \xdocref{(18)}.

\begin{prop}\label{prop:oracle_mean}
    Suppose that the conditions in Assumption~\xdocref{1} hold for each $X_k$, $k=1,\dots,K$. Then, under the condition that $\vv_{\textsc{j}} = \Vv_{(k)} \wv_{(k)}$ for some weights $\wv_{(k)}$ and all $k \in [K]$, we have
    \begin{align}\label{eq:oracle_mean_1}
        \langle  \vv_{\textsc{j}} , \tilde{\vv}_{(k)}^\ast \rangle &= 
        m_{(k)} +O_p(n^{-1/2}),
    \end{align}
    where $m_{(k)} = \sum_{i=1}^{r_k}w_{(k),i}w^\circ_{(k),i} m_{y_k}(\lambda_{(k),i}) = 1/\|\Dv_{(k)}^{-1}\wv_{(k)}\|_2$.
    Furthermore, for any $\epsilon >0$, we have
    \begin{align}\label{eq:oracle_mean_2}
        \Eb  \langle \vv_{\textsc{j}} , \tilde{\vv}_{(k)}^\ast \rangle = m_{(k)} + o(n^{-1/2+\epsilon}).
    \end{align}
    Consequently, $ T_\ast^2 = m^\circ + O_p(n^{-1/2})$ where $m^\circ = K+2 \sum_{k<l} m_{(k)} m_{(l)}$.
\end{prop}

Proposition~\ref{prop:oracle_mean} justifies the form of $m_{(k)} = \sum_{i=1}^{r_k} w_{(k),i}w_{(k),i}^\circ m_{y_k}(\lambda_{(k),i})$, the limiting mean of $\langle \vv_{\textsc{j}}, \tilde{\vv}^\ast_{(k)} \rangle$, as well as our estimator $\hat{m}_{(k)}$ in \xdocref{(18)}. 
By Theorem~\xdocref{5}, this further yields the limiting mean of $T^2_\ast$ as $m^\circ = K+2 \sum_{k<l} m_{(k)} m_{(l)}$ and estimate it by $\tilde{\mu}_\yv(\widehat{\lambdav})$ in \xdocref{(17)}.
The limiting variance of $\langle \vv_{\textsc{j}}, \tilde{\vv}^\ast_{(k)} \rangle$, however, is difficult to evaluate, so we approximate it by ignoring the cross-product terms in \xdocref{(20)}.
We therefore focus on the variance of the diagonal term in \xdocref{(20)},
\begin{align*}
    A_{(k)}:= \sum_{i=1}^{r_k} w_{(k),i} w_{(k),i}^\circ \langle \vv_{(k),i}, \hat{\vv}_{(k),i}\rangle,
\end{align*}
which is a weighted sum of the $r_k$ terms $\langle \vv_{(k),i}, \hat{\vv}_{(k),i}\rangle$.

By \citet{bao2021singular}, we have that for any bounded continuous function $f$,
\begin{equation}\label{eq:asymp_equiv}
\begin{aligned}
    &\lim_{n\to \infty}\left[\Eb f \left(\sqrt{n}\left\{\langle \vv_{(k),i}, \hat{\vv}_{(k),i} \rangle^2 - m^2_{y_k}(\lambda_{(k),i})\right\} \right) - \Eb f(\Zc_{(k),i}) \right] =0 \quad \textnormal{for all } i \in [r_k], \\
    & \lim_{n\to \infty}\left[\Eb f \left(\sqrt{n}\left\{ \sum_{i=1}^{r_k} \langle \vv_{(k),i}, \hat{\vv}_{(k),i} \rangle^2 - \sum_{i=1}^{r_k}m^2_{y_k(\lambda_{(k),i})}\right\} \right) - \Eb f\left(\sum_{i=1}^{r_k}\Zc_{(k),i} \right) \right] =0
\end{aligned}
\end{equation}
where $\Zc_{(k),i}\overset{\mathrm{ind}}{\sim}\Nc\!\big(0,\Vc^\circ_{y_k}(\lambda_{(k),i})\big)$ for $i=1,\dots,r_k$.
\eqref{eq:asymp_equiv} implies that 
\begin{align*}
    &\sqrt{n}\left\{\langle \vv_{(k),i}, \hat{\vv}_{(k),i} \rangle^2 - m^2_{y_k}(\lambda_{(k),i})\right\} \simeq \Zc_{(k),i}, \quad \textnormal{and} \\
    &\sqrt{n}\left\{ \sum_{i=1}^{r_k} \langle \vv_{(k),i}, \hat{\vv}_{(k),i} \rangle^2 - \sum_{i=1}^{r_k}m^2_{y_k(\lambda_{(k),i})}\right\} \simeq \sum_{i=1}^{r_k}\Zc_{(k),i},
\end{align*}
where $\simeq$ denotes asymptotic equivalence in distribution, i.e., both sides have the same limiting law. 
Thus, at the $\sqrt n$ scale, the centered alignment quantities can be approximated by the corresponding Gaussian limits for variance calculations.
Since the Gaussian limits $\{\Zc_{(k),i}\}_{i=1}^{r_k}$ are mutually independent, the cross-covariance terms across different spike indices vanish, motivating an independence approximation when aggregating fluctuations across $i$.
Under this working approximation, we estimate the variance of $A_{(k)}$ by
$\sum_{i=1}^{r_k} \Var \left( w_{(k),i} w^\circ_{(k),i} \langle \vv_{(k),i}, \hat{\vv}_{(k),i} \rangle \right) = \sum_{i=1}^{r_k} (w_{(k),i} w_{(k),i}^{\circ} )^2\,\Var\big(\langle \vv_{(k),i}, \hat{\vv}_{(k),i} \rangle\big)$.
Incorporating the plug-in parameter estimates and estimated coefficient vector $\hat{\wv}_{(k)}=\hat{\Vv}_{(k)}^\top\hat{\vv}_{\textsc{j}}/\|\hat{\Vv}_{(k)}^\top\hat{\vv}_{\textsc{j}}\|$ and $\hat{\wv}_{(k)}^\circ = \hat{\Dv_k}^{-1}\hat{\wv}_{(k)}/\|\hat{\Dv_k}^{-1}\hat{\wv}_{(k)}\|_2$, we approximate the asymptotic $n$-scaled variance of $\langle \vv_{\textsc{j}}, \tilde{\vv}^\ast_{(k)} \rangle$ by $\tilde{\Vc}_{(k)}$ in \xdocref{(18)} and estimate it by $\tilde{\Vc}_\yv^\circ(\widehat{\lambdav})$ in \xdocref{(17)}.

\begin{proof}[Proof of Proposition~\ref{prop:oracle_mean}]
    By the definition of $\vv_{\textsc{j}}$ and $\tilde{\vv}^\ast_{(k)}$, we can write
    \begin{align*}
        \langle \vv_{\textsc{j}} , \tilde{\vv}_{(k)}^\ast \rangle 
        =& \langle \Vv_{(k)}\wv_{(k)} , \hat{\Vv}_{(k)} \wv_{(k)}^\circ \rangle\\
        =&  \sum_{i=1}^{r_k} w_{(k),i}w_{(k),i}^\circ \langle \hat{\vv}_{(k),i}, {\vv}_{(k),i} \rangle + \sum_{i<j}^{r_k} w_{(k),i}w_{(k),j}^\circ \langle \hat{\vv}_{(k),i}, {\vv}_{(k),j} \rangle \\
        =& A_{(k)} + B_{(k)},
    \end{align*}
    where $A_{(k)}:=\sum_{i=1}^{r_k} w_{(k),i}w_{(k),i}^\circ \langle \hat{\vv}_{(k),i}, {\vv}_{(k),i} \rangle$ and $B_{(k)} = \sum_{i<j}^{r_k} w_{(k),i}w_{(k),j}^\circ \langle \hat{\vv}_{(k),i}, {\vv}_{(k),j} \rangle$.
    By Lemma~\ref{thm:asym.dist.angles.Bao}, $A_{(k)}=m_{(k)}+O_p(n^{-1/2})$ and $B_{(k)}=O_p(n^{-1/2})$, which claims the first argument~\eqref{eq:oracle_mean_1}.
    
    For the second argument~\eqref{eq:oracle_mean_2}, first note that each $\mathbb E\langle \vv_{(k),i},\hat{\vv}_{(k),j}\rangle=0$ for all $i \neq j \in [r_k]$.
    This can be checked by using a simple symmetry argument.
    Reflecting both the $j$-th left and right signal singular directions of signal matrix leaves $\Av_k$ unchanged, while $\Ev_k$ has the same distribution. 
    Hence $\Xv_k$ is distributionally unchanged. 
    % With the sign convention $\langle \hat{\vv}_{(k),l},\vv_{(k),l}\rangle\ge0$,
    For $i\neq j$, this reflection preserves the orientation of $\hat{\vv}_{(k),i}$ relative to $\vv_{(k),i}$, but changes the sign of $\langle \vv_{(k),i},\hat{\vv}_{(k),j}\rangle$. Therefore $\langle \vv_{(k),i},\hat{\vv}_{(k),j}\rangle\stackrel{d}{=}-\langle \vv_{(k),i},\hat{\vv}_{(k),j}\rangle$, and consequently $\mathbb E\langle \vv_{(k),i},\hat{\vv}_{(k),j}\rangle=0$.
    As a result, $\sqrt{n}\, \Eb \, B_{(k)} = 0$.

    By \citet{ding2020high}, for any small $\epsilon>0$ and large $D>0$, the following holds with probability at least $1 - n^{-D}$ for sufficiently large $n$:
    \begin{align*}
        \left| \langle \vv_{(k),i}, \hat{\vv}_{(k),i} \rangle^2 - m^2_{y_k}(\lambda_{(k),i}) \right| \leq n^{-\frac{1}{2}+\frac{\epsilon}{2}} \left[ \frac{1}{ \delta^{1/2} }  +  \frac{1}{n^{1/2} \delta^{4} } \right].
    \end{align*}
    Under the supercritical condition, $m_{y_k}(\lambda_{(k),i})$ is uniformly bounded away from zero. Hence, for large enough $n$,
    \begin{align*}
        \left| \langle \vv_{(k),i}, \hat{\vv}_{(k),i} \rangle - m_{y_k}(\lambda_{(k),i}) \right| \leq C n^{-\frac{1}{2}+\frac{\epsilon}{2}},
    \end{align*}
    for some universal constant $C$, which claim that $\Eb \, A_{(k)} = m_{(k)}+o(n^{-\frac{1}{2}+\epsilon})$.
\end{proof}

\bibliographystyle{asa}
\bibliography{mssat}